\documentclass[11pt,a4paper, showpreprintnumber]{article}
\pdfoutput=1
\usepackage{jheppub_r}
\usepackage{graphicx}
\usepackage{epsfig}
\usepackage{amsmath, amssymb}
\usepackage{dcolumn}
\usepackage{bm}
\usepackage{amsfonts}
\usepackage{epstopdf}
\usepackage{enumerate}

\newcommand{\nc}{\newcommand}

\newcommand{\hf}{\frac{1}{2}}
\newcommand{\bea}[1]{\begin{eqnarray} \mbox{$\label{#1}$}}
\newcommand{\eea}{\end{eqnarray}}
\newcommand{\be}[1]{\begin{equation} \mbox{$\label{#1}$}}
\newcommand{\ee}{\vspace{0.1cm}\end{equation}}
\newcommand{\eq}[1]{\mbox{(\ref{#1})}}
\newcommand{\fig}[1]{\mbox{Fig.\ (\ref{#1})}}
\newcommand{\sect}[1]{\mbox{section\ \ref{#1}}}
\newcommand{\nl}{\nonumber \\}
\def\GeV{{\rm \ GeV}}

\nc{\fb}[2]{\left(\frac{#1}{#2}\right)}
\nc{\sqb}[2]{\sqrt{\frac{#1}{#2}}}

\nc{\jeq}{j_{_{\rm EQ}}}
\nc{\jnp}{j_{_{\rm NP}}}
\nc{\jbr}{j_{_{\rm BR}}}
\nc{\jnr}{j_{_{\rm NR}}}
\nc{\gT}{g_{_{\rm T}}}

\newcommand{\qnr}{q_{\mathrm{nr}}}
\newcommand{\knr}{k_{\mathrm{nr}}}
\newcommand{\anr}{a_{\mathrm{nr}}}
\newcommand{\znr}{z_{\mathrm{nr}}}
\newcommand{\tnr}{t_{\mathrm{nr}}}

\nc{\mT}{m(T)}
\nc{\mTsq}{m^2(T)}
\nc{\mTstar}{m(T_*)}
\nc{\mTstarsq}{m^2(T_*)}

\nc{\msq}{m^2_\sigma}

\nc{\Mpl}{M_{\mathrm{P}}}
\nc{\sigstar}{\sigma_*}
\nc{\msig}{m_{\sigma}}
\nc{\ud}{\text{d}}

\nc{\kcut}{k_{\mathrm{c}}}
\nc{\kmax}{k_{\mathrm{max}}}

\nc{\fnl}{f_{_{NL}}}

\newcommand{\Xphi}{X_{\phi}}

\title{Reheating dynamics affects non-perturbative decay of spectator fields}
\author{Kari Enqvist,}
\emailAdd{kari.enqvist@helsinki.fi}
\author[1]{Rose N. Lerner\note{From October 2013, rose.lerner@desy.de.}}
\emailAdd{rose.lerner@helsinki.fi}
\author{and Stanislav Rusak}
\emailAdd{stanislav.rusak@helsinki.fi}
\affiliation{University of Helsinki and Helsinki Institute of Physics, P.O. Box 64, FI-00014, Helsinki, Finland.}

\preprint{HIP-2013-11/TH}


\abstract{The behaviour of oscillating scalar spectator fields after inflation depends on the thermal background produced by inflaton decay. Resonant decay of the spectator is often blocked by large induced thermal masses. We account for the finite decay width of the inflaton and the protracted build-up of the thermal bath to determine the early evolution of a homogeneous spectator field $\sigma$ coupled to the Higgs Boson $\Phi$ through the term $g^2 \sigma^2 \Phi^2$, the only renormalisable coupling of a new scalar to the Standard Model. We find that for very large higgs-spectator coupling $g\gtrsim 10^{-3}$, the resonance is {\em not} always blocked as was previously suggested. As a consequence, the oscillating spectator can decay quickly. For other parameter values, we find that although qualitative features of the thermal blocking still hold, the dynamics are altered compared to the instant decay case. These findings are important for curvaton models, where the oscillating field must be relatively long lived in order to produce the curvature perturbation. They are also relevant for other spectator fields, which must decay sufficiently early to avoid spoiling the predictions of baryogenesis and nucleosynthesis.}

\begin{document}
\maketitle

\section{Introduction}\label{intro}

Inflation may or may not be driven by a light scalar field, but there exists at least one light scalar spectator, a field that plays no dynamical role during inflation but is subject to almost scale invariant fluctuations. This is the Higgs field\footnote{There have been many attempts to generate inflation with the Higgs field by tuning its coupling to gravity, see \cite{higgsinflation}.}, but there could be others. A much studied example is the curvaton \cite{curvaton}, which is a spectator that well after the end of inflation becomes dynamically important and through its late decay gives rise to the observed curvature perturbation.  Another example is a spectator that couples to the inflaton and whose perturbations induce fluctuations in the inflaton decay rate, giving rise to modulated (p)reheating and adiabatic perturbations \cite{modulated}. (For a recent discussion on Planck implications for spectator models, see \cite{Kobayashi:2013bna}.) There has also been some discussion about the role of the Higgs
field in the generation of the curvature perturbation \cite{higgspert,Enqvist:2013kaa}.

Spectators could also be harmful. Light scalar spectators with non-zero expectation values are in slow roll during inflation. After inflation ends, they can remain in slow roll for some time. However, the Hubble parameter $H(t)$ is decreasing as the universe expands. Thus, the light fields will eventually become heavy, $m \gg H(t)$, and start to oscillate. Meanwhile, the energy density that drove inflation is being converted into radiation with an effective decay width $\Gamma$, in the simplest case by the decay of the scalar inflaton field to massless fields, which then thermalise. Inflaton decay can take place either before or after the onset of spectator oscillations. For a weakly coupled field this oscillation can continue for some time, and a spectator field that has not decayed could come to dominate the Universe, generate gravitational waves, affect the process of baryogenesis, or even spoil the predictions of big bang nucleosynthesis (BBN). A related issue is whether the spectator
field could be dark matter. This is discussed for our model in \cite{Cline:2013gha}.

The cosmological consequences of spectators thus depend on the time (and sometimes the specific process) by which the oscillating field decays. Thus, it is important to understand in detail the mechanism by which spectator fields decay. Here we do not specify the nature of inflation but merely assume that there is an inflaton-like decay and that the decay products thermalise fast. To be specific, we assume that the thermal bath so generated consists of the Standard Model (SM) particles, and that the spectator fields are coupled to the thermal bath. We also assume that the spectators are SM singlets. Then the only renormalizable coupling of the spectator $\sigma$ to the SM is to the Higgs field. This can happen through  either  a three-point or a four-point interaction. The former case would require a mass parameter to control the spectator decay rate. Instead, we impose the discrete symmetry $\sigma\to -\sigma$ so that the only remaining possibility is a coupling of the type
$g^2\sigma^2\Phi^\dagger\Phi$,
where $\Phi$ is the Higgs doublet and $g$ a (small) coupling constant. Thus, after inflaton decay, there is a thermal bath of higgses into which the spectator $\sigma$ couples to. Although this set-up is specific, it could easily be generalised to other scalar couplings.

There are three main mechanisms by which oscillating fields can ``decay''\footnote{In this context, we use ``decay'' to mean the process by which the homogeneous, oscillating field transfers its energy into (thermalised) particles}. The first is perturbative decay. At tree-level, this requires a three-point coupling in the Lagrangian, which we do not have. There is also a perturbative effect due to the presence of the thermal bath. This is effective only for large spectator-higgs couplings, and is discussed further in \sect{pert}. The other effects are both non-perturbative decay processes. {\em Broad resonance} can occur for some sets of parameters and is a collective field theory effect in which non-adiabatic evolution while the oscillating field crosses zero causes particle production. In contrast, {\em narrow resonance} can be understood as a two-to-two body process by which the oscillating field generates quanta of its decay product, although it is still a non-perturbative process. Broad
resonance is typically fast and efficient, whereas narrow resonance can take longer to complete.

It was recently shown that the presence of a thermal background coupled to the oscillating field's decay product can completely {\em block} both broad resonance and narrow resonance while the background temperature is large enough \cite{Enqvist:2012tc}. This is because the thermal background gives a large effective mass to the fields. This blocking is then lifted as the temperature falls, eventually allowing decay. For some parameters, the decay appears to be extremely inefficient and may not complete at this point.

Thus, the thermal background tends to substantially delay the ``decay'' of the spectator field, making viable curvaton models that would otherwise be ruled out due to a too-fast decay \cite{Enqvist:2012tc}. However, \cite{Enqvist:2012tc} relied on the assumption that the inflaton field decayed into thermalised radiation instantly, before the first opportunity for resonant decay occurred. This is not necessarily a reasonable assumption, and it is the aim of this paper to study the effect of a realistic production of a thermal background on the decay process of an oscillating spectator field. Therefore, we investigate the dynamics of the spectator field and the resonant particle production as the inflaton decay rate is varied. The fate of spectator fields, and the curvaton in particular, depends on the time of their decay, and thus on whether non-perturbative decay occurs.

The topic of the perturbative dissipation of a scalar in a thermal background has been studied in the context of warm inflation \cite{BasteroGil:warminflation}. Related work \cite{Mukaida:2012qn} discusses thermal effects on an oscillating scalar field coupled to a generic fermion, covering the dissipative effect of the thermal bath, modification of the scalar potential, the formation of non-topological solitons and non-perturbative particle production.

The remainder of the paper is organised as follows. In \sect{dynamics}, we set up the dynamics of the model. We also give an overview of the parameter space, which contains three main cases. In \sect{inflaton}, we discuss the dynamics of inflaton decay and calculate the background temperature. In sections \ref{resgT}-\ref{resCW} we calculate the time of onset and the efficiency of the non-perturbative decay in each of these three cases. In \sect{pert}, we provide a calculation of the perturbative decay due to the thermal bath. In \sect{results},  we collect and combine the results of the previous sections, presenting the parameter space where spectator decay occurs before inflaton decay. Finally, \sect{conc} contains our discussion and conclusion.

\section{Spectator dynamics}\label{dynamics}

Our semi-realistic model consists of only the SM, one real singlet scalar spectator field $\sigma$ coupled to the higgs, and some mechanism for inflation. The inflaton is only specified through its effective decay width to the SM, given by $\Gamma$. It need not be a scalar inflaton, and if so, need not be coupled to the spectator\footnote{Once a specific mechanism for inflaton is specified, additional constraints may apply to the model.}. Thus, inflation need not necessarily be of particle physics origin, as long as the production of SM particles can be modelled by some effective rate $\Gamma$. We assume a relativistic fluid and an instant thermalisation for inflaton decay products. Our scenario is motivated because it is a simple extension of the SM where all renormalisable couplings of the spectator and the SM are included. The Lagrangian is
\be{lag}
{\cal L} = {\cal L}_{\mathrm{SM}} + {\cal L}_{\mathrm{inf}} + \hf \partial_\mu \sigma \partial^\mu \sigma +  \frac{1}{2}m^2_\sigma\sigma^2 + \lambda \sigma^4 + g^2\sigma^2\Phi^\dag\Phi,
\ee
where ${\cal L}_{\mathrm{SM}}$ and ${\cal L}_{\mathrm{inf}}$ contain the SM and inflaton sectors respectively. For simplicity, we now set\footnote{For curvaton models, the effect of including $\lambda \neq 0$ and higher dimensional operators has been widely considered \cite{selfinteracting}.} $\lambda = 0$, although in \sect{resCW} we do consider the radiatively-generated four-point function.  The relevant parameters of the model are the value of the spectator field at the end of inflation $\sigma_*$, the ``bare'' mass of the spectator $m_\sigma$, the effective decay width of the inflaton $\Gamma$, the spectator-higgs coupling $g$ and the Hubble parameter at the end of inflation $H_*$. We assume that after inflation ends, the inflaton energy density can be described by an oscillating scalar field in a quadratic potential, thus giving an effectively matter-dominated Universe (until inflaton decay).

An important feature of the model is that both the higgs and the spectator gain (potentially very large) effective masses in the presence of any thermal background of SM particles. The effective scalar masses become
\be{meffH}
m^2_{h}(T) \simeq (126\GeV)^2 + g^2  \sigma^2  + \gT^2 T^2
\ee
and
\be{meffsigma}
m^2_{\sigma}(T) \simeq m^2_\sigma + g^2 T^2,
\ee
where the effective thermal coupling of the higgs $\gT^2 = 0.1$ is known \cite{Anderson:1991zb}. These approximations for the thermal masses require the particles running in the loop to be relativistic; this is always true for the SM particles contributing to the higgs mass, but is only true for the higgs contributing to the spectator's mass if $g^2\sigma^2 \ll T^2$. If the condition is not met, the spectator's thermal correction in \eq{meffsigma} is exponentially suppressed.

Quantum corrections can also be important for the spectator. These result in an effective one-loop correction to the spectator's potential given by
the Coleman-Weinberg term
\bea{oneloop}
\label{VCW}
 V_{1 \mathrm{loop}} \simeq \frac{g^4\sigma^4}{64\pi^2}\left[\ln\left(\frac{g^2\sigma^2}{m^2}\right) - \frac{3}{2}\right],
\eea
which is valid in the limit $g\sigma \gg T$ (non-relativistic higgs). In the opposite limit, the correction is suppressed and anyway subdominant compared to the thermal mass of the spectator. This one loop correction can dominate the potential during inflation for large $\sigma_*$, small $m_\sigma$ and large $g$.

The equation of motion of the spectator and the higgs (components suppressed) are given by
\be{eomsigma}
\ddot \sigma + 3H \dot \sigma + (m_{\sigma}^2 + g^2T^2)\sigma = 0
\ee
and
\be{eomH}
\ddot \phi + 3H\dot \phi + \left(\frac{k^2}{a^2} + g^2\sigma^2  + \gT^2T^2\right)\phi = 0,
\ee
where $\phi$ is the expectation value of $\Phi$.

The spectator should be light compared to the Hubble parameter during inflation for our calculations to apply, otherwise it will roll to the minimum of its potential. Note that during inflation both the higgs and the spectator are light fields, and gain a spectrum of perturbations. Because of their coupling, their mean field values contribute to the effective mass of both fields. The mean field value of the higgs is approximately $\overline\phi\equiv\sqrt{\langle \phi_*^2\rangle}\simeq 0.36 \lambda^{-1/4} H_{*}\simeq 1.1 H_*$, derived using stochastic principles\footnote{Because of the relatively large higgs self-coupling, the Higgs field attains its equilibrium distribution in a few tens of $e$-folds.} (see e.g.\ \cite{Enqvist:2012xn}). Thus, $g H_*$ provides a lower limit to the effective mass of the spectator during inflation. For all values of $g$ that we consider, the spectator field remains light during inflation. The effective mass during inflation may be relevant to calculate predictions for the curvaton model, but
not to determine the decay process of a general spectator. We assume that after the end of inflation, due to its large couplings, the higgs decays well before the inflaton (see \cite{Enqvist:2013kaa} for discussion of the Higgs condensate during inflation).

\begin{figure}
\begin{center}
\includegraphics[width=0.47\textwidth]{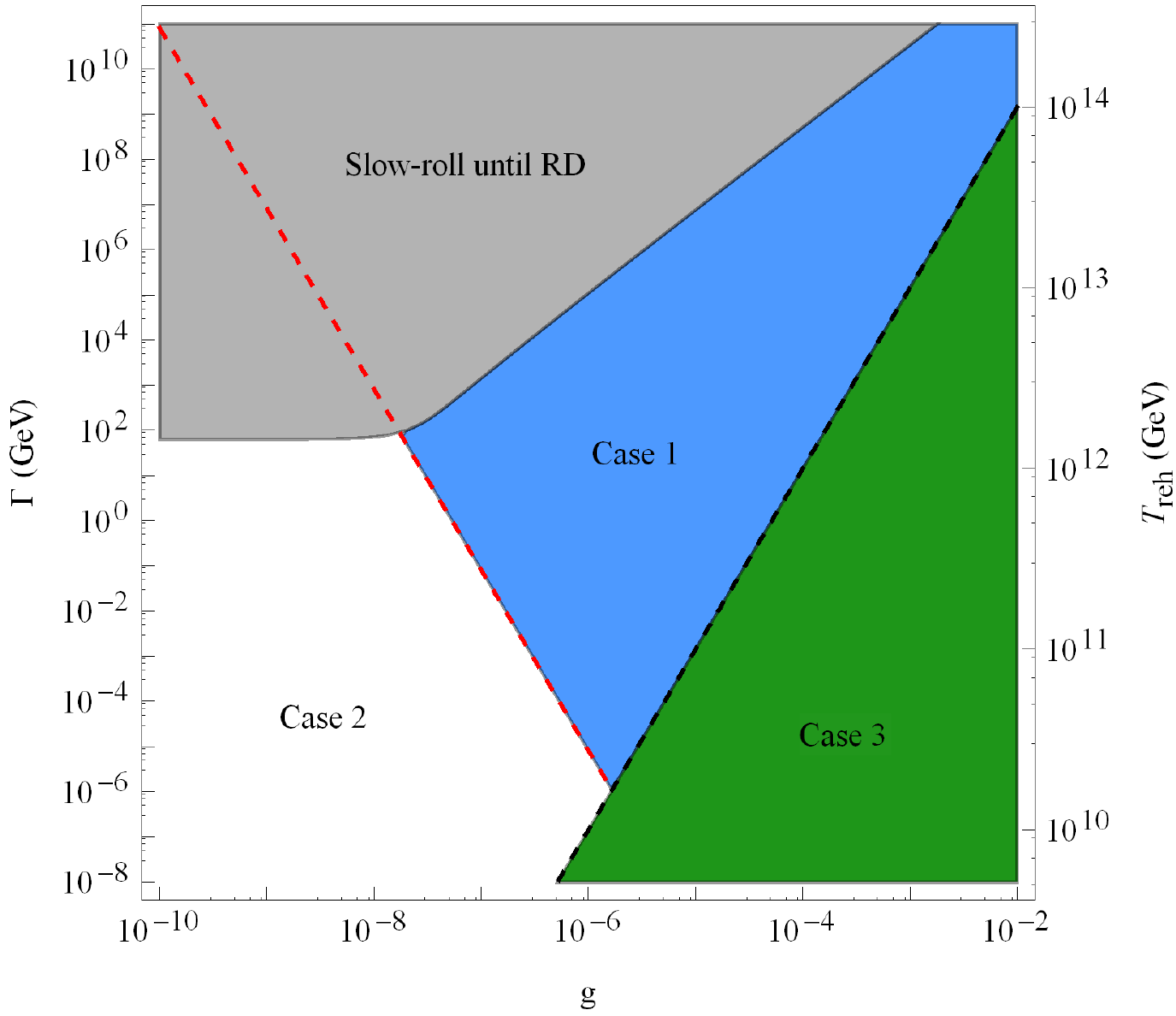} \qquad
\includegraphics[width=0.47\textwidth]{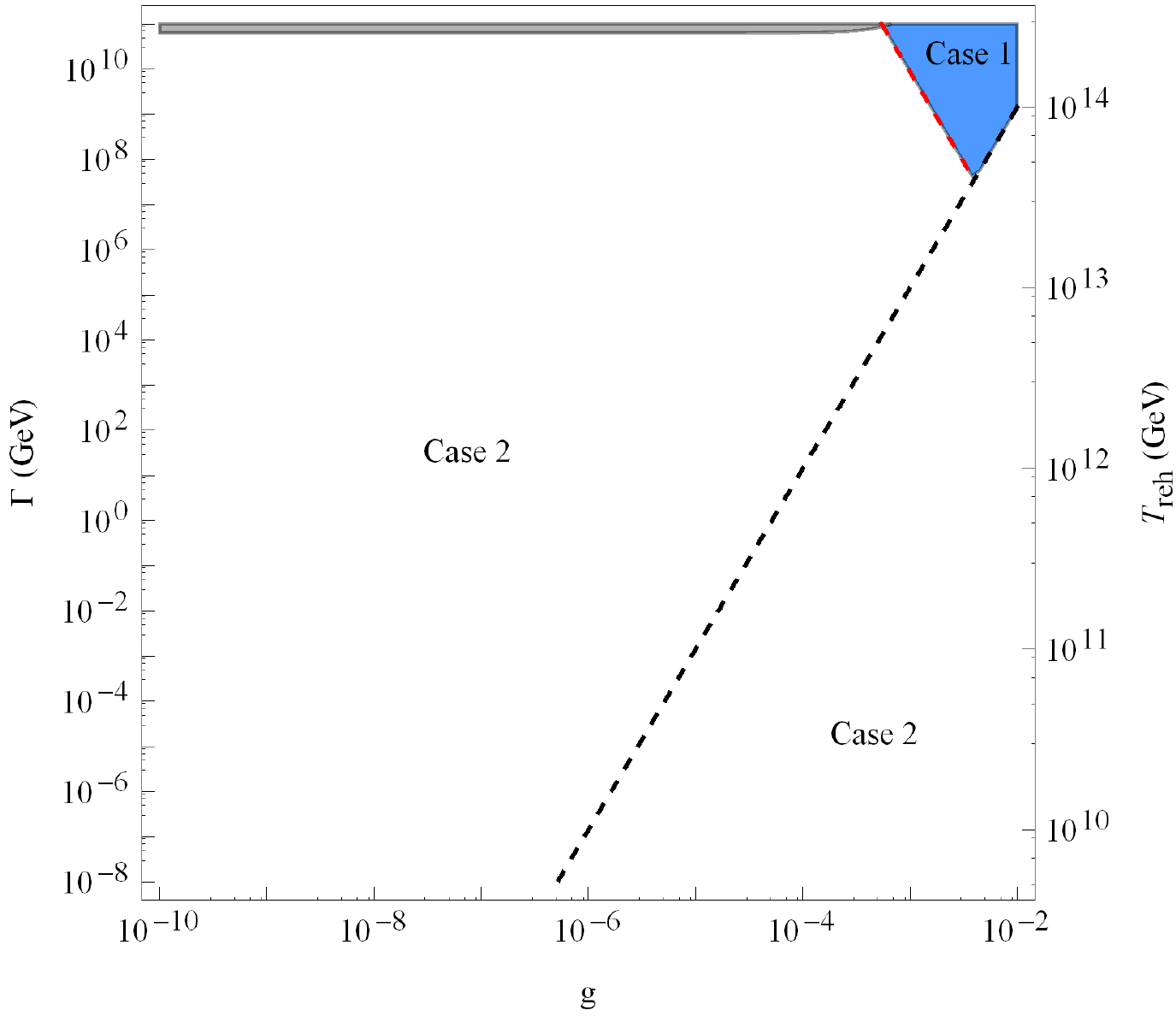}
\caption{\label{fig:region}Regions of parameter space distinguished by their evolution at the first zero-crossing, shown for both small ($m_\sigma = 100\GeV$, left) and large ($m_\sigma = 10^{11}\GeV$, right) spectator masses. Case 1 (blue) has $m_\sigma(T) = gT$, case 2 (white) has $m_\sigma(T) = m_\sigma$, case 3 (green) has the effective potential dominated by the one-loop term, and in the light grey region the spectator does not oscillate before inflaton decay is complete. The red dashed line shows $m_\sigma = g T$ at the first zero-crossing. Below the black dashed line, the higgs is initially non-relativistic. ($H_* = 10^{12}\GeV$.)}
\end{center}
\end{figure}

After inflation, the spectator field oscillates (for $H(t) < m_\sigma(t)$) with decaying amplitude. The effective frequency is set by the effective mass, which could be dominated by $m_\sigma$, the temperature correction, or the 1-loop potential. In order to calculate whether the spectator decays non-perturbatively, we first need to know the background dynamics of the system, and specifically when the spectator crosses zero. These are dependent on the model parameters. In our analytical approximations, this divides the parameter space into a number of regions (\fig{fig:region}), which depend on the system at the first zero crossing of the spectator.

Case 1 is shown in blue --- here the spectator's effective mass is dominated by its thermal correction, and the first zero crossing occurs between the maximum of the background temperature and the end of inflaton decay. Case 2 is shown in white --- here the spectator's effective mass is dominated instead by $m_\sigma$, and again the first zero crossing occurs between the maximum of the background temperature and the end of inflaton decay. Case 3 is shown in green --- here the one-loop corrections dominate the potential at the end of inflation. To the right of the black dashed line, the higgs is initially non-relativistic and therefore the spectator cannot obtain a thermal mass for any $m_\sigma$. The red line shows $gT = m_\sigma$ at the first zero-crossing. In the grey region, the first crossing occurs {\em after} the inflaton has decayed and the radiation-dominated period has begun\footnote{In this case, the spectator is in slow roll while the inflaton is decaying, and to a first approximation, its amplitude does not change. Thus, the
results of \cite{Enqvist:2012tc} are valid, but with the substitutions $T_* \to T_{\mathrm{eq}}$ and $t_0 \to t_{\mathrm{eq}}$, where ``eq'' refers to the matter-radiation equality at $H(t)\sim \Gamma$.}. It is cases 1, 2 and 3 that are the focus of this paper. In the following three sections, we obtain analytical solutions for $\sigma(t)$, and for the onset and efficiency of the resonance in these three regions. Varying $H_*$ hardly changes \fig{fig:region}, except that for large $g$ the spectator can become heavy during inflation. Decreasing $\sigma_*$ increases the region of case 1 and decreases the region of case 3.

We cannot usefully scan over all parameters, so we present results for $m_\sigma = 100 \GeV$ and $m_\sigma = 0.1H_*$. We fix $H_* = 10^{12}\GeV$ and $\sigma_* = 10^{16}\GeV$ for the figures, and discuss the dependence of our results on these parameters.

\section{Inflaton decay}\label{inflaton}

Although inflaton decay and generation of a thermal background is assumed to be instantaneous in previous studies of curvaton reheating (e.g.\ \cite{Enqvist:2012tc}), this is an obvious simplification. The particular relevance of the timescale of the generation of the SM thermal background arises because the existence of a thermal background can cause blocking of an otherwise-efficient resonance \cite{Enqvist:2012tc}. However, if the timescale to generate the thermal background is very long, this blocking could in principle be prevented. In that case, the spectator would be likely to decay very quickly before the thermal background is established. This outcome is `helpful' for an unwanted spectator, but `unhelpful' for a curvaton-spectator. This is not a minor alteration of the predictions, but a completely different outcome for the model. In addition, even if the resonant decay is still blocked, a long period of effective matter-domination (before inflaton decay) alters the expansion history of the Universe,
 and will require a re-calculation of predictions.

We are interested in slow decay of the inflaton, $\Gamma \ll H_*$, where $\Gamma$ is the effective decay rate of the inflaton. Various requirements give a lower limit for $\Gamma$. For example, a strict lower limit comes from the requirement that the inflaton decays before big bang nucleosynthesis occurs, restricting $\Gamma \gtrsim 10^{-22} \GeV$. Baryogenesis and dark matter production would in all likelihood yield much more stringent lower limits, but as these are model-dependent, we do not include them. We vary $\Gamma$ from $10^{-8}\GeV$ to $10^{11}\GeV$; this range demonstrates the main behaviours of the model.

Our focus is on the period between the end of inflation and the decay of the inflaton. During this period, we assume the Universe is effectively matter dominated, and $3H^2(t)\Mpl^2 \propto a^{-3}$, where the scale factor is
\be{aMD}
a = \left(1+3H_*t/2\right)^{2/3}.
\ee
We model inflaton decay and the production of the thermalised SM particles with
\be{g2}
 \dot \rho_{\mathrm{inf}} + 3H\rho_{\mathrm{inf}}  = - \Gamma \rho_{\mathrm{inf}}
\ee
 and
\be{g_1}
 \dot \rho_{\mathrm{SM}} + 4H\rho_{\mathrm{SM}}  = \Gamma \rho_{\mathrm{inf}}.
\ee
The thermal energy density in SM particles as the inflaton decays is given by \bea{rhoSM}
 \rho_{\mathrm{SM}} & = & 3\Mpl^2H_*^2\left(\frac{a}{a_0}\right)^{-4}\int_0^{t-t_0}d t' \left(\frac{a}{a_0}\right) \Gamma e^{-\Gamma t'} \nl
& \simeq & \frac{6}{5}\Mpl^2H_* \Gamma a^{-4}\left[a^{5/2}e^{-\Gamma t} -1\right]\left[1+\mathcal O\left(\frac{\Gamma}{H_*}\right)\right].
\eea
This rises rapidly at first, then falls approximately $\propto a^{-3/2}$ until the inflaton has decayed, when it then falls $\propto a^{-4}$. The period where $\rho_{\mathrm{SM}} \propto a^{-3/2}$ is due to the competing effects of the expansion of the  Universe and the production of new particles. Assuming instant thermalisation after decay, the temperature is then given by the usual relation
\be{rhoSM_T}
\rho_{\mathrm{SM}}=\frac{\pi^2}{30}g_*T^4,
\ee
where $g_*=106.75$ counts the SM degrees of freedom. We then find the time-temperature relationship
\be{fulltemp}
T(t)
\simeq \left[\frac{36}{\pi^2 g_*} \frac{\Mpl^2 H_* \Gamma}{(1+3H_* t/2)}\right]^{1/4}.
\ee
The reheating temperature is defined as the maximum temperature; an analytical estimate gives
\be{T_max}
T_{\mathrm{reh}} \simeq 0.3 (\Mpl^2H_* \Gamma)^{1/4}.
\ee
We also define an asymptotic temperature $T_*$ such that for $T<T_{\mathrm{reh}}$, the temperature is given by
\be{asymptoticT}
T(t) = T_* a^{-3/8}(t).
\ee

\begin{figure}[tb]
\begin{center}
\includegraphics[width=0.6\textwidth]{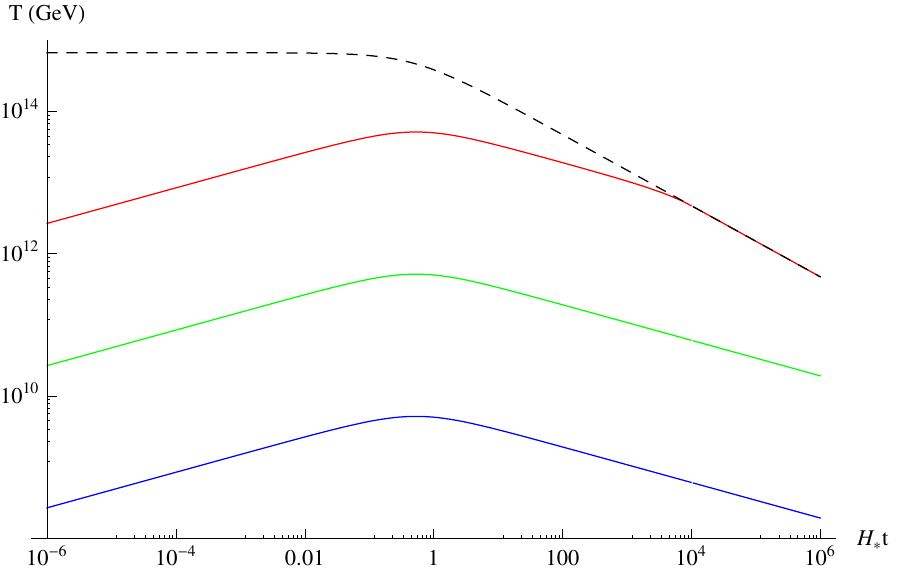}
\caption{Temperature of the SM background generated by inflaton decay. From top to bottom: instant decay approximation (dashed), $\Gamma = 10^{-4} H_* = 10^8\GeV$ (red), $\Gamma = 10^{-12} H_* = 1\GeV$ (green) and $\Gamma = 10^{-20}H_* = 10^{-8}\GeV$ (blue). Note that the reheating temperature is reached well before reheating is complete, which occurs at $t \sim 1/\Gamma \gg 1/H_*$ ($H_* = 10^{12}\GeV$).\label{fig:temp}}
\end{center}
\end{figure}
\fig{fig:temp} shows the SM temperature as a function of time for $\Gamma = 10^{-4} H_* = 10^8\GeV$ (second from top, red), $\Gamma = 10^{-12} H_* = 1\GeV$ (middle, green) and $\Gamma = 10^{-20}H_* = 10^{-8}\GeV$ (bottom, blue). The instant decay approximation is also shown (top, dashed). Important features to note are that the temperature peaks when only a small fraction of inflaton has decayed, i.e.\ at $t_{\mathrm{max}}\ll 1/\Gamma$. Second, after the maximum the temperature decreases slower than $a^{-1}$, because of the continuing inflaton decay. Third, the reheating temperature is reduced compared to the instant decay approximation.

As mentioned before, we assume instant thermalisation of the inflaton's decay products. This is fully justified when $m_\sigma(T) = \gT T$: the timescales for thermalisation and the oscillation period of the spectator are $t_{\mathrm{osc}} \sim 1/(g T)$ and $t_{\mathrm{therm}} \sim 1/T$. Given that $g <1 $, we find $t_{\mathrm{osc}} > t_{\mathrm{therm}}$. If instead $m_\sigma(T) = m_\sigma$, then using $t_{\mathrm{osc}} = 1/m_\sigma$ we see that only for smaller $m_\sigma$ is the instant thermalisation approximation valid. However, the issue is not really whether there is thermal equilibrium, but instead whether the timescales of particle production allow the produced particle to ``see'' the background{\footnote{Only kinetic (not chemical) equilibrium is required for our conclusions to hold. In the case of a non-thermal bath of relativistic particles, the spectator and the higgs masses would still receive finite density corrections of the same order as the thermal mass (if the forward scattering rate in the plasma was sufficiently fast).}}. For the
remainder of this paper, we assume that the instant thermalisation approximation is sufficient.

For clarification, we emphasise that immediately after inflation, the Universe is dominated by the oscillating inflaton field, for convenience assumed here to be in a quadratic potential such as the universe behaves as matter dominated, and $H(t) \propto a^{-3/2}$. Then when the inflaton has decayed, the Universe is effectively radiation-dominated, such that $H(t)\propto a^{-2}$. At some later time it is possible for the {\em spectator} to dominate the Universe, oscillating in a quadratic potential. This would give rise to a second matter-dominated period. After spectator decay, there would then be a second radiation-dominated period. Finally, as the Universe cools, the usual matter-dominated period will occur (well after both spectator and inflaton have decayed).

\section{Case 1: spectator mass is thermal}\label{resgT}
The calculations in this section apply to the blue region of the parameter space (\fig{fig:region}), where the thermal correction dominates the spectator's effective mass with $m_\sigma\ll gT$.

\subsection{Background dynamics}
The equation of motion of the spectator \eq{eomsigma} can be solved analytically when $\rho_{\mathrm{SM}} \propto a^{-3/2}$ (see \eq{rhoSM}).
Substituting for $H(t)$ and $T(t)$ gives
\be{eomsigma_9}
\ddot \sigma + 3\frac{H_*}{(1+3H_* t/2)} \dot \sigma + \left[m_{\sigma}^2 + g^2\left(\frac{36}{\pi^2 g_*} \frac{\Mpl^2 H_* \Gamma}{(1+3H_* t/2)}\right)^{1/2}\right]\sigma = 0.
\ee
In the limit $m_\sigma = 0$, this has the general solution
\be{gensol}
\sigma = C_1 \, a^{-3/4}(t)\,\, J_{\frac{2}{3}}\left(A a^{9/8}(t)\right) + C_2 \, a^{-3/4}(t) \, \, J_{-\frac{2}{3}}\left(A a^{9/8}(t)\right)
\ee
where
\be{Adef}
 A \simeq  \frac{1.3}{g_*^{1/4}} g\sqrt{\frac{\Mpl}{H_*}}\left(\frac{\Gamma}{H_*}\right)^{\frac{1}{4}}.
\ee
Matching to the initial condition $\sigma(0) = \sigma_*$ gives
\bea{sigMD}
 \sigma(t)  & \simeq & \Sigma(t) \sin\left(Aa^{9/8} - \frac{\pi}{12}\right),
\eea
where
\bea{Sigma}
\Sigma(t) & \equiv & \frac{\sigma_*}{A^{7/6}} \frac{1}{a^{21/16}(t)}.
\eea
The resonant production of higgs particles depends on the zero-crossings of the spectator, which occur for integer values of $j$ at
\be{t_j}
t_j \simeq 2.3 g_*^{1/3} \left(g^4\Mpl^2 \Gamma\right)^{-\frac{1}{3}}\left(j + \frac{1}{12}\right)^{\frac{4}{3}}, \qquad j = 1,2,3,\dots \, .
\ee
This solution is valid provided that $t > t_{\mathrm{reh}}$, i.e.\ $\rho_{\mathrm{SM}} \propto a^{-3/2}$, and $t\leq 1/\Gamma$ i.e.\ spectator decay is not yet complete.

\subsection{Broad Resonance}
\label{sec:gT}
Non-perturbative decay occurs by broad resonance (rather than narrow), provided that the resonance parameter $q$ is large,
\be{q_def}
q(t) \equiv \fb{g\Sigma(t)}{2m(t)}^2 \gg 1.
\ee
We find $q \gg 1$ at $T = T_*$ in case 1, provided that $\sigma_* \gtrsim T_*$. Note that narrow resonance is always blocked in this case because $g^2 T^2 \ll \gT^2 T^2$ (see \sect{narrow}).

During broad resonance, particles are produced only in a short time interval around each zero crossing of the spectator. We denote the time of each zero crossing as $t_j$, and expand $\sigma(t)$ around $t_j$. Thus, for $t = t_j + \delta t$ with $\delta t \ll (t_{j+1} - t_j)$, we find
\be{approx_sigma}
\sigma(t_j + \delta t) \simeq \frac{1.3}{A^{1/6}} \frac{\sigma_* H_*}{a_j^{27/16}} \delta t.
\ee
We emphasise that $t_j$ is a fixed constant for each zero crossing, and the only time-dependence is in $\delta t$.
Thus, the linearised equation of motion for the higgs is
\bea{eomH_lin}
\ddot \phi + 3H\dot \phi + \left(\frac{k^2}{a_j^2} + \frac{1.6}{A^{1/3}} \frac{g^2 \sigma_*^2 H_*^2}{a_j^{27/8}} \delta t^2  + \gT^2T_j^2\right)\phi &=& 0.
\eea

For ease of calculation, we write \eq{eomH_lin} as
\be{eom_final}
\frac{d^2 \chi}{d\tau^2} + \left(\kappa_j^2 + \tau^2\right)\chi,
\ee
where the variables have been transformed as follows, and small corrections to the effective frequency have been neglected. The field is written as
\be{change}
\chi = a^{3/2}(t) \phi,
\ee
the new time variable is defined as
\bea{def_tau}
\tau & = & \frac{\kcut}{a_j} \delta t,
\eea
and
\bea{def_kappa}
\kappa_j^2 & \simeq & \fb{k}{\kcut}^2 + \fb{a_j^{5/8} \gT T_*}{\kcut}^2,
\eea
where
\be{kcut}
\kcut \equiv \frac{\sqrt{g\sigma_* H_*}}{A^{1/12}} a_j^{5/32}
\ee
defines the cut-off scale below which the momenta evolve non-adiabatically {\em in the absence of thermal corrections}.

Broad resonance requires non-adiabatic evolution in some interval around the zero crossing. This occurs if $\frac{d\omega}{d\tau} \gg \omega^2$, where $\omega$ is the effective frequency of the  {\em higgs} oscillation, given by
\be{omega}
\omega^2 = \kappa_j^2 + \tau^2.
\ee
Thus, adibaticity is violated for a brief interval $\Delta \tau \lesssim 1$ if
\be{cond_1}
 0 \leq k^2 \leq \frac{2}{3\sqrt{3}} \kcut^2 - a_j^{5/4} \gT^2 T_*^2
\ee
is satisfied. Note the sign of the term containing the temperature --- it works to block the resonance. This can give a  situation where no values of $k$ that satisfy \eq{cond_1} exist. Indeed, we find that in this region of parameter space (case 1, where the spectator mass is thermal), \eq{cond_1} is never satisfied. Thus, the main conclusion of this section is that for case 1, the resonance is always blocked until late times.

\section{Case 2: bare spectator mass dominates}
\label{resm}
The calculations in this section apply to the white region of the parameter space (\fig{fig:region}), where $m_\sigma$ dominates the spectator's effective mass with $m_\sigma\gg gT$. In this case, it is necessary to consider both broad and narrow resonance.

\subsection{Background Dynamics}
 In this case, the spectator sees the matter-like background and
\bea{u2}
\sigma(t) & = & \frac{\sigma_*}{a^{3/2}(t)}\left[ \cos m_\sigma t + \left(\frac{3 H_*}{2m_\sigma} + \frac{\dot\sigma_*}{m_\sigma\sigma_*}\right)\sin m_\sigma t\right]
 \\
& \simeq & \Sigma(t) \sin\left(m_\sigma t + \frac{2 m_\sigma}{3H_*} \right),
\eea
where
\be{signewagain}
\Sigma(t) \equiv \frac{3 H_* \sigma_*}{2 m_\sigma\,a^{3/2}(t)},
\ee
valid for $m_\sigma t > \pi/2$.
If the field is initially in slow-roll, the zero crossings are at $t_j \simeq j\pi/m_\sigma$ with $j = 1, 2, 3,...$ The first zero crossing is before inflaton decay is complete ($H(t)\sim\Gamma$) provided that $\pi \Gamma /m_\sigma<1$.

\subsection{Broad Resonance}

In this case, the spectator does not gain a large thermal mass from the higgs, either because the higgs is non-relativistic, or because the magnitude of $gT$ is much smaller than $m_\sigma$. Note that the SM particles {\em are} still relativistic, so the higgs does still gain the thermal mass $\gT T$. If the resonance parameter $q>1$ (see \eq{q_def}), we consider broad resonance as follows.

The transformed equation of motion \eq{eom_final} remains valid, as do the definitions of $\kappa_j$ and $\tau$. However, $\kcut$ is different and is now given by
\be{newkcut}
\kcut \equiv \sqb{3H_* g \sigma_*}{2}a_j^{1/4}.
\ee
The resonance again occurs for modes which satisfy
\be{cond_p}
 0 \leq k^2 \leq \left(k_{\mathrm{max}}^j\right)^2 \equiv \pi^{-1}\kcut^2 - \gT^2T_*^2a_j^{5/4}.
\ee
We find that in some fraction of the parameter space where this solution is valid, the resonance is initially unblocked. This occurs for large $m_\sigma$, large $g$ and small $\Gamma$ (see \fig{fig:case2}). Although $\kcut$ is different compared to case 1, the main contribution to the unblocking of the resonance is that the case 2 parameter space has larger $g$ and smaller $\Gamma$ than the case 1 parameter space.

An important novelty in this situation is that, in contrast to the scenario presented in \cite{Enqvist:2012tc}, the temperature correction {\em increases} its contribution to the blocking as time increases\footnote{This is valid for $H(t) \lesssim \Gamma$, i.e.\ while the inflaton is decaying; in the radiation-dominated era the relative contribution will fall again.}. Thus, if the resonance is initially {\em unblocked}, it becomes blocked after $j_{\mathrm{block}}$ zero-crossings,
\begin{equation}
j_{\mathrm{block}} = \frac{g_*}{24\pi \gT^4} g^2 \bigg(\frac{\sigma_*}{\Mpl}\bigg)^2\bigg(\frac{m_\sigma}{\Gamma}\bigg).
\end{equation}
Note however that the resonance may have terminated prior to $j_{\mathrm{block}}$ if it transferred all of the energy out of the spectator.  We have checked that the resonance always becomes blocked\footnote{Note that in the radiation-dominated era, the blocking term again decreases in magnitude, and we expect the resonance to become unblocked a second time as in \cite{Enqvist:2012tc}.} (if it has not already completed) before inflaton decay is complete at $H(t)\sim\Gamma$. This is an interesting result and different from the calculation in the instant inflaton decay approximation.

If non-perturbative decay is unblocked, it is important to calculate whether it has time to efficiently transfer energy out of the spectator to the SM before the resonance becomes blocked. We do this in the following section.

\subsubsection{Efficiency of particle production}
\label{sec:effcase2}
Still considering broad resonance in case 2, we calculate the decay efficiency. To determine this, we need to calculate the energy transferred to the higgs. The spectator energy density after $j$ oscillations is
\be{rho_sig}
 \rho_{\sigma} \simeq \frac{1}{2}\msig^2\Sigma_j^2 = \frac{\msig^2\sigstar^2}{2\pi^2j^2}
\ee
The higgs occupation number after the $j$th oscillation is \cite{Kofman:1997yn}
\bea{nk_full}
 n_k^{j+1} & = & e^{-\pi \kappa_j^2} + \left(1+2e^{-\pi \kappa_j^2}\right)n_k^j - 2e^{-\pi \kappa_j^2/2}\sqrt{1+e^{-\pi \kappa_j^2}}\sqrt{n_k^j(1+n_k^j)}\sin\theta_{\mathrm{tot}}^j
 \nl
 & \simeq & \left(1+2e^{-\pi \kappa_j^2}  - 2 \sin\theta_{\mathrm{tot}}^j e^{-\pi \kappa_j^2/2}\sqrt{1+e^{-\pi \kappa_j^2}}\right)n_k^j \nl
 & \equiv & e^{2\pi \mu_k^j} n_k^j
\eea
where $\theta_{\mathrm{tot}}^j$ is the total phase, and the final lines are valid for large occupation number. The phase behaves stochastically, so the $\sin\theta_{\mathrm{tot}}^j$ term should average to zero over many oscillations. In our case, there are no particles before the first zero-crossing because the higgs is non-relativistic. Therefore, \eq{nk_full} reduces to
\be{nk}
n_k^{j+1} \simeq \frac{1}{2}e^{2\pi\sum_{i=1}^j \mu_k^{i}}.
\ee
Modes are excited if $\pi \kappa_j^2 < 1$. As noted earlier, the temperature correction causes the resonance band to become narrower with each oscillation, in contrast to the traditional preheating scenario.

Our analytical work does not include backreaction of the produced higgs particles on the spectator. Therefore, we follow standard procedure and deem that the resonance has efficiently completed when $\rho_\sigma = \rho_{\mathrm{h}}$. Using the fact that only modes with $|k| \leq \kmax^j$ are amplified\footnote{Note that although $\kmax$ decreases with $j$, the exponential amplification ensures that using an integration limit of $\kmax^j$ is sufficient for our accuracy.} and that the produced higgses are non-relativistic, the energy density in the higgs at the $(j+1)^{th}$ zero-crossing is given by
\be{eq:rho_h_from_sigma}
 \rho_{\mathrm{h}}^{j+1} \simeq  \frac{4 m_{h,j+1}}{2\pi^2a_{j+1}^3}\int_0^{\kmax^j} \ud k k^2 e^{2\pi\sum_1^j \mu_k^j} n_k^1.
\ee

Following the procedure in Appendix A, we approximate the integral in \eq{eq:rho_h_from_sigma} and obtain
\bea{finalrho_h}
 \rho_{\mathrm{h}}  &\simeq &\frac{1}{4\pi^{11/2}}\frac{(g\msig\sigstar)^{3/2}g\sigstar \,\, \exp\left[\left(\ln 3 - \frac{4}{9}\sqrt{j/j_{\mathrm{block}}} + \frac{1}{18}j/j_{\mathrm{block}}\right)j\right]  }{j^4 \left(1 - \frac{4}{21}\sqrt{j/j_{\mathrm{block}}}\right)^{3/2}}.
\eea
As we are only interested in whether the decay is efficient, and not the precise moment when it completes, we calculate $\rho_h / \rho_\sigma$ at $j = j_{\mathrm{block}}$. If the ratio is larger than 1, we declare the resonant decay to have been efficient. We find that in some of the region where the resonance is initially unblocked, it is also efficient and completes before becoming blocked.

As a final remark, one might be concerned that the higgs becomes temporarily relativistic near the spectator's zero-crossing, and consequently that the non-relativistic assumption is not valid. We have checked that this is not the case, by comparing timescales of non-adiabaticity and thermalisation of the higgs in the thermal bath, $\Delta t_{\mathrm{therm}} \sim 1/(\gT T)$. We find that there is not enough time for the thermal  bath to generate higgs particles before the resonance, and thus that our results do not need adjusting.

\subsection{Narrow resonance}
\label{narrow}
If the resonance parameter  is small, $q<1$ (see \eq{q_def}), then broad resonance as described above cannot occur. Instead, narrow resonance is a possibility. In this section we determine whether this occurs.

Narrow resonance can be thought of as a $2\to 2$ process, where two zero-mode spectator quanta produce two higgs quanta, each with energy $E_k$. Particle production occurs when (following \cite{Enqvist:2012tc})
\be{narrow_1}
2m_\sigma = 2E_k = 2\left( \frac{k^2}{a^2} + 4q(t)m_\sigma^2\sin^2\left(m_\sigma t \right) + \gT^2 T^2\right).
\ee
To determine the limiting condition, we consider the $k = 0$ mode, and recall that $ q < 1$. Thus, particle production can only occur if
\be{g1}
m_\sigma > \gT T.
\ee
The dynamics of particle production can be described by the Mathieu equation \cite{Kofman:1997yn} whose solutions are amplified $\propto e^{\mu_k\msig t}$ (see Appendix~\ref{sec:appendixB}) within narrow instability regions (where the narrow resonance Floquet index $\mu_k$ is real and positive). Due to the redshifting of momenta, modes will enter the resonance band, where they get amplified, and then exit again after a short time. The time spent inside the resonance band is $\Delta t \sim q t$, where the $t$ is the time when the mode enters. Because $q$ decreases with time and because higher $k$-modes enter the resonance band later, these higher $k$-modes spend a shorter time within the band than the lower $k$-modes. Eventually the production stops because the resonance band becomes too narrow for modes to be significantly amplified. The magnitude of $\mu_k$ is also proportional to $q$ and therefore
$\mu_k \msig \Delta t \propto q^2\msig t = \msig q_{\mathrm{nr}}^2 \tnr(a/a_{\mathrm{nr}})^{-9/2}$ where the subscript `nr' refers to the time of the onset of narrow resonance.
In order to have significant particle production we must have
\be{hhh1}
q_{\mathrm{nr}}^2\msig \tnr = q_*^2(\msig \tnr)^{-1}\gg1.
\ee
There are two ways for narrow resonance to begin. If we are initially in the narrow resonance regime ($q<1$) then particle production starts once it becomes thermally unblocked, that is,
when the condition~\eqref{g1} is satisfied. In this case IR modes are excited (see Appendix~\ref{appendixB2}). The time of the onset is given by
\begin{equation}
 \msig \tnr \simeq \frac{24}{g_*\pi^2}\left(\frac{\Mpl}{\msig}\right)^2\left(\frac{\Gamma}{\msig}\right),
\end{equation}
calculated using \eq{hhh1}. We find that in this case the condition for significant production
\begin{equation}
 q_{\mathrm{nr}}^2\msig \tnr = \frac{g_*\pi^2}{384}g^4\left(\frac{\sigstar}{\Mpl}\right)^2\left(\frac{\sigstar}{\msig}\right)^2\left(\frac{\msig}{\Gamma}\right) > 1
\end{equation}
is almost never satisfied so the spectator will not decay efficiently by initial narrow resonance.

The second way for narrow resonance to begin is if the transition from broad resonance happens long after the condition~\eqref{g1} becomes satisfied. In this case the time of onset is given by $q\sim 1$,
\begin{equation}
 \msig \tnr \simeq \sqrt{q_*} = \frac{g}{2}\left(\frac{\sigstar}{\msig}\right).
\end{equation}
In contrast to the previous case, UV modes are amplified and the energy density is (see Appendix~\ref{appendixB3})
\begin{eqnarray}
 \rho_h \simeq  \frac{8\msig^4}{27\pi^2}\frac{e^{\frac{3}{2}\qnr^2\msig \tnr\left(1-\beta\right)}}{\qnr^2\msig\tnr\left(1+2\beta/9\right)}\left(\frac{a}{\anr}\right)^{-4}, \label{eq:NR_energy_density}
\end{eqnarray}
where $\beta \simeq (3\qnr^2\tnr T_{\mathrm{nr}}/2)^{-1}$. Significant production requires both that $\qnr^2\msig \tnr$ is large and that $\beta$ is small; without the latter
there are too few particles initially for efficient production to occur. As in the case of broad resonance, we declare that the spectator decays efficiently if $\rho_h \sim \rho_{\sigma}$.

\begin{figure}
\centering
\includegraphics[width=0.6\textwidth]{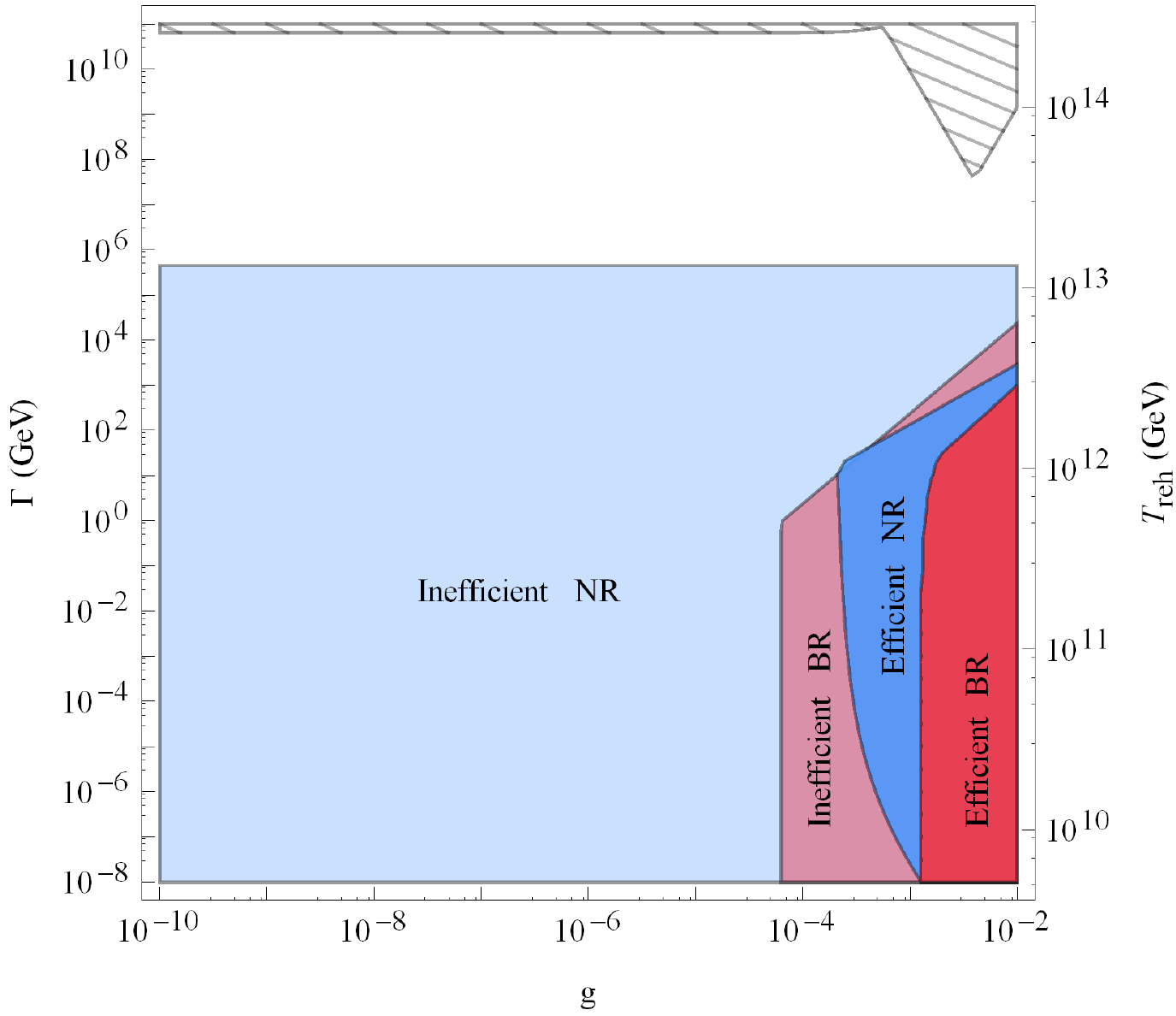}
\caption{\label{fig:case2} Case 2: regions where non-perturbative decay occurs before the inflaton has decayed. We show only $m_\sigma = 10^{11}\GeV$; for $m_\sigma = 100\GeV$ there is no resonance. In the dark red and dark blue regions, the resonance is efficient and the spectator {\em decays} before $H(t)\geq \Gamma$. The dark red region is broad resonance; the dark blue region is narrow resonance. Broad resonance begins but is not efficient for smaller $g$, encompassing both the light red region, and the dark blue region where narrow resonance {\em is} efficient. Inefficient narrow resonance occurs in much of the remaining parameter space (light blue). In the hashed region, case 2 is not valid. Plotted with $H_* = 10^{12}\GeV$ and $\sigma_* = 10^{16}\GeV$.}
\end{figure}

For $m_\sigma = 100\GeV$, $H_* = 10^{12}\GeV$ and $\sigma_* = 10^{16}\GeV$ we find no region of parameter space where the spectator decays before the inflaton in case 2. At larger $m_\sigma$, both broad and narrow resonance become possible. \fig{fig:case2} shows where the non-perturbative decay occurs before $H(t)\sim\Gamma$. As expected, this requires smaller $\Gamma$ corresponding to a lower reheating temperature. Broad resonance is only efficient for the largest values of $g$ (dark red). Narrow resonance takes over and is efficient in the dark blue region. The light red region and the dark blue region (smallest $g$) have inefficient broad resonance. There is also inefficient narrow resonance, which occurs in much of the remaining parameter space (light blue).

The exact shapes of \fig{fig:case2} depend on the approximations made; however the general conclusion holds, that for large $g$ and large $m_\sigma$, if the inflaton decays slowly, the thermal blocking may be prevented.

\section{Case 3: Coleman-Weinberg correction dominates the potential}
\label{resCW}
The calculations in this section apply to the green region of the parameter space (\fig{fig:region}), where $V_{1 \mathrm{loop}}$, given by (\ref{VCW}), dominates the spectator's effective potential during inflation.

\subsection{Background Dynamics}
We approximate the one-loop potential by $V \simeq \frac{g^4}{64\pi^2}\sigma^4$. The background spectator solution is given by
\be{sigmaCW}
 \sigma \simeq \left\{\begin{array}{lcr}
    \sigstar & \quad \mathrm{for} & a(t) < 1.2/{\tilde{\lambda}}^{1/3}
                    \\
    \frac{\sigstar}{a(t)} \left(\frac{1.2}{ {\tilde{\lambda}}^{1/3} }\right) \mathrm{cn}\left[2.76\left(0.9\, \tilde{\lambda}^{1/6}\, a^{1/2}(t)  -1\right),\frac{1}{\sqrt{2}}\right] & \quad \mathrm{for} & a(t) > 1.2/{\tilde{\lambda}}^{1/3}
                  \end{array} \right. ,
\ee
where $\mathrm{cn}(x-x_0,1/\sqrt{2})$ is an elliptic cosine function, and
\be{tildelam}
\tilde \lambda = \frac{g^4}{16\pi^2}\left(\frac{\sigstar}{H_*}\right)^2.
\ee
This is different from the usual massless preheating case because of the period of effective matter domination, which delays the first oscillation of the spectator.

The elliptic cosine is periodic with the period given by $\approx 1.85$.
Thus, the spectator crosses zero at
\be{CWzero}
a_j = \frac{1.2}{\tilde{\lambda}^{1/3}}\left[1+ 0.67(2j-1)\right]^2, \qquad j = 1,2,3,\cdots.
\ee

\subsection{Broad Resonance}

The model in this case corresponds to the massless preheating model~\cite{Greene:1997fu}, but with complications due to the matter-dominated background and the thermal background. In principle, both spectator quanta and higgs quanta can be excited at the zero crossings although the production of higgs quanta is expected to be dominant \cite{Greene:1997fu}. Defining $X_\phi \equiv a^{3/2} \phi_k$ and linearising around the spectator's zero crossings gives
\begin{equation}
\frac{d \Xphi}{d\tau^2} + \left(\kappa_j^2 + \tau^2\right)\Xphi = 0,
\end{equation}
where
\be{xcw}
\tau \equiv \frac{k_c}{a_j}\delta t,
\ee
\be{kccw}
k_c \equiv \frac{g \sigma_* H_*}{\tilde{\lambda}^{1/6}}\delta t,
\ee
and
\begin{equation}
 \kappa_j^2 \equiv \fb{k}{k_c}^2 + \frac{\gT^2 T_*^2 a_j^{5/4}}{k_c^2}.
\end{equation}
The resonance occurs if $\pi\kappa_j^2<1$, which is true for modes that satisfy
\be{eq:resonance_band_sig5}
 k^2 \leq k_{\mathrm{max}}^2 \equiv \frac{k_c^2}{\pi} - \gT^2 T_*^2 a_j^{5/4}.
\ee
As in case 2, the scaling of $T\propto a^{-3/8}$ means that the resonance will eventually become blocked, if it hasn't already completed by that point. This happens at
\begin{eqnarray}
 j_{\mathrm{block}} & = & 0.67\fb{k_c^2}{\pi \gT^2 T_*^2}^{2/5}
 \\
 & \simeq & 0.5 \left(\frac{g}{\gT}\right)^{4/5}\left(\frac{\sigstar}{\Mpl}\right)^{3/5}\left(\frac{\Mpl}{\Gamma}\right)^{1/5}.
\end{eqnarray}

\subsubsection{Efficiency of the broad resonance}
Still considering case 3, we now calculate the proportion of that region where the resonance is efficient. The calculation follows that for case 2, found in \sect{sec:effcase2}. Using the fact that the produced higgses are non-relativistic and ignoring the phase factor{\footnote{The phase does not behave stochastically as in case 2 because the situation is almost-conformal. This means that the maximum of the resonance band will not be at $k=0$ but will instead be determined by the coupling. When considering the whole parameter space, we believe this effect averages out.}}, the higgs energy density is
\begin{equation}
 \rho_{\mathrm{h}}^{j+1} \simeq \frac{4m_{h,j}}{2\pi^2 a_j^2}\int_0^{k_{\mathrm{max}}^j}\ud k k^2 \frac{1}{2}e^{2\pi\mu} \simeq \frac{5 g\sigstar e^{2\pi \mu_0}}{16\pi^2 \tilde{\lambda}^{1/3}a_j^3 |\mu_0''|^{3/2}}.
 \label{eq:rho_h_from_sigma_sig4}
\end{equation}
where $\mu_0$ and $\mu''_0$ are given by \eq{am0} and \eq{amp} in the Appendix.

\fig{fig:case3} shows the regions of parameter space where the resonance occurs and is efficient in case 3 for $m_\sigma = 100\GeV$. The resonance occurs only for small $\Gamma$ and large $g$, and decay completes before $H(t)\sim\Gamma$ in some of the parameter space (dark red). For very large $m_\sigma$, the 1-loop potential never dominates, so case 3 is never relevant.
\begin{figure}
\centering
\includegraphics[width=0.6\textwidth]{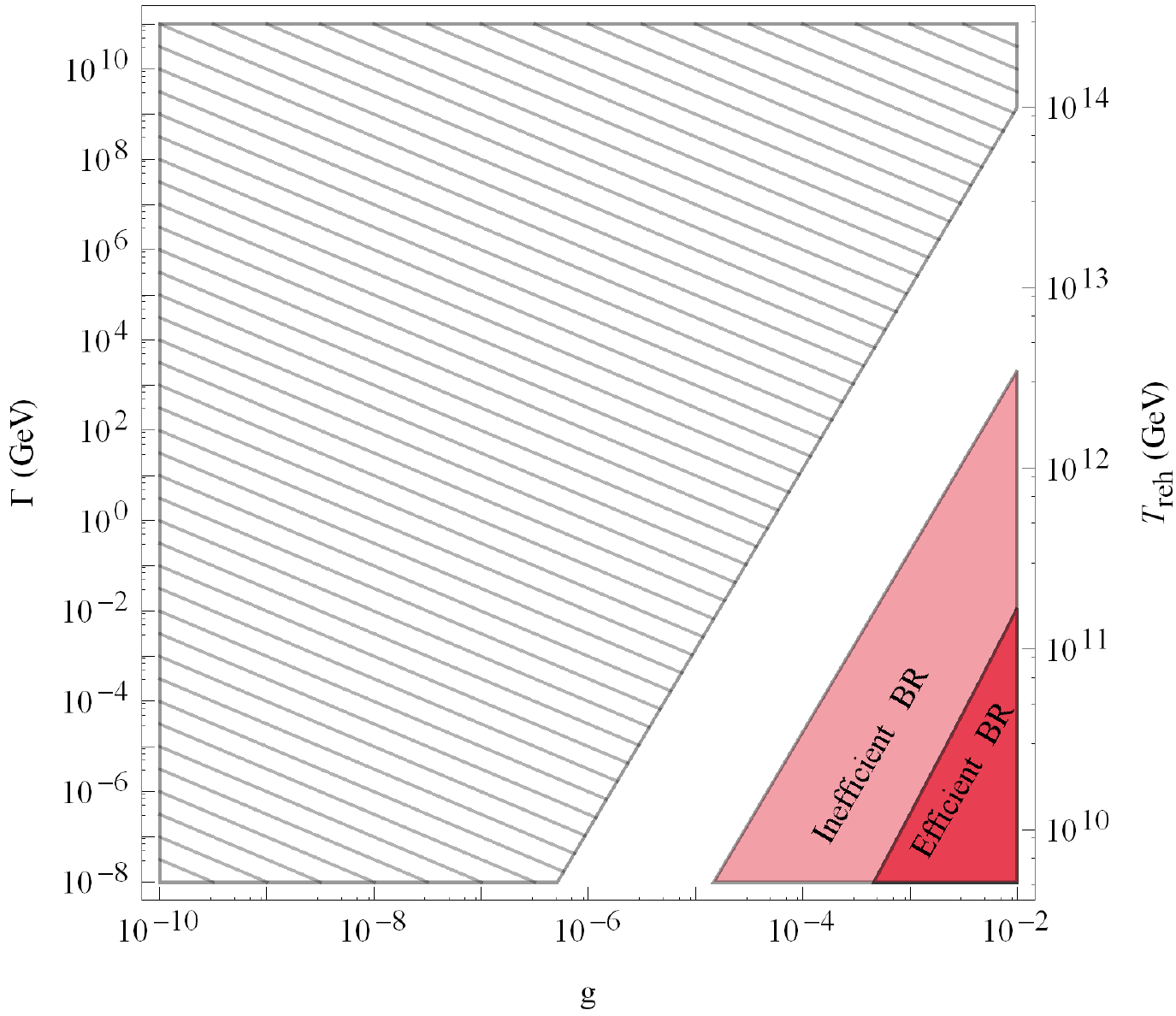}
\caption{\label{fig:case3} Case 3: regions where non-perturbative decay occurs before the inflaton has decayed. We show only $m_\sigma = 100\GeV$; for $m_\sigma = 10^{11}\GeV$ there is no region where case 3 is valid. In the shaded region (red), non-perturbative decay of the spectator by broad resonance begins before $H(t)\geq \Gamma$. In the dark red region the resonance is efficient and the spectator decays before the inflaton. In the hashed region, case 3 is not valid. Plotted with $H_* = 10^{12}\GeV$ and $\sigma_* = 10^{16}\GeV$.}
\end{figure}

\section{Perturbative decay due to thermal bath}\label{pert}
In addition to the non-perturbative processes already discussed, the presence of the thermal bath means that additional decay processes for the spectator can be relevant, including scatterings with the thermal background. The full thermal decay rate $\Gamma_{\mathrm{th}}$ of the scalar field zero mode, which includes both scattering,  $1\to 3$ particle decays and inverse decays, is related to the imaginary part of the self-energy ${\rm Im}\Sigma$ by $\Gamma_{\mathrm{th}}=-{\rm Im}\Sigma/m(T)$. In thermal field theory, to lowest order the self-energy is given by a two loop diagram, the so-called sunset diagram, and has been calculated in \cite{Elmfors:1993re}.  For cases 1 and 2, the decay rate is
 \be{new1}
\Gamma_{\mathrm{th}} = \frac{1}{576\pi} \frac{g^4 T^2}{m_\sigma(T)},
\ee
valid provided that both $m_\sigma(T) \ll T$ and $m_h\simeq g\Sigma(t) \ll T$. If this condition is not fulfilled, then there is exponential suppression of the decay rate.
For the 1-loop potential \eq{oneloop}, the rate is
\be{extranew1}
\Gamma_{\mathrm{th}} = 6.6\times 10^{-8} \frac{g^8 T^2}{m_\sigma(T)},
\ee
which is also only valid for $m_\sigma(T) \ll T$. The decay rates depend on time; therefore we determine the time of decay by calculating when $\Gamma_{\mathrm{th}} = H(t)$.

In case 1 ($m_\sigma(T) = gT$), decay occurs when
\be{gamma1}
 H(t) = 1.5\times 10^{-5} g^4 \Mpl^{2/3}\Gamma^{1/3}.
\ee
In this case, $m(T) = gT < T$ is always satisfied. For $H_* = 10^{12}\GeV$ and $\sigma_* = 10^{16}\GeV$, there is no perturbative decay before $H(t)\sim\Gamma$ in this case, except for large $g$ and small $m_\sigma$.

In case 2 ($m_\sigma(T) = m_\sigma$), decay occurs when
\be{gamma2}
 H(t) = 10^{-8} \frac{g^8 \Mpl^{2}\Gamma}{m_\sigma^2},
\ee provided the curvaton and higgs are both relativistic. For $H_* = 10^{12}\GeV$ and $\sigma_* = 10^{16}\GeV$, there is no perturbative decay before $H=\Gamma$ in this case.

In case 3 ($V(\sigma) = V_{1 \mathrm{loop}}(\sigma)$), there are two contributions. The quartic potential (from \eq{extranew1}) gives
\be{gamma3}
 H(t) = 10^{-16} \frac{g^{16} \Mpl^{2}\Gamma}{m_\sigma^2},
\ee
again provided that the condition $m_\sigma < T$ is satisfied. This expression describes processes only involving the spectator. However, because $m_\sigma(T) = m_\sigma$ for case 3, the coupling to the higgs then gives \eq{gamma2}, which dominates over \eq{gamma3}. In the case where the higgs is non-relativistic but the spectator is relativistic, only \eq{gamma3} is valid. However, this rate is always less than the inflaton decay rate $\Gamma$.

\begin{figure}
\centering
\includegraphics[width=0.60\textwidth]{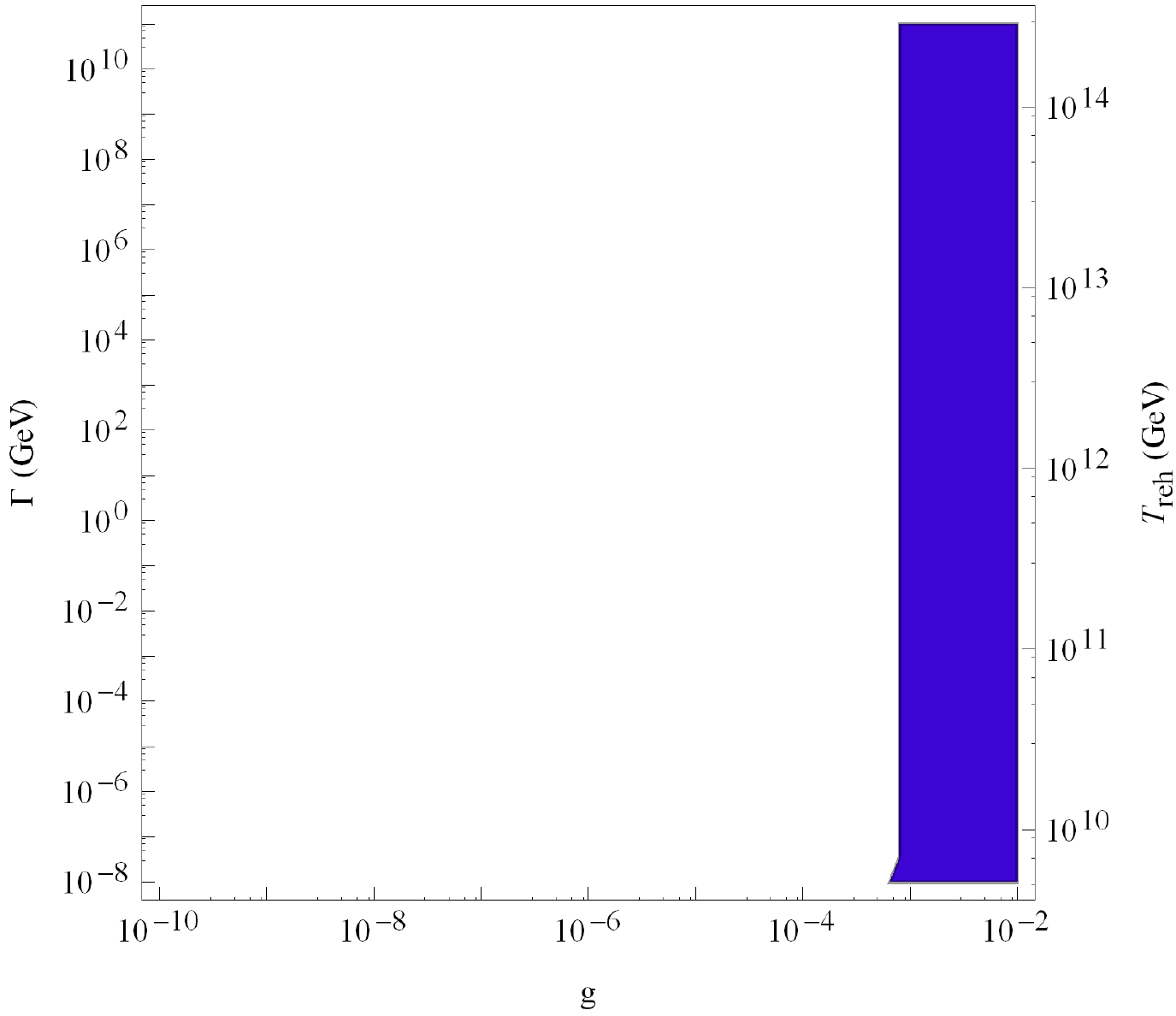}
\caption{\label{fig:pert} Shaded region shows where perturbative decay in the presence of the thermal bath occurs before inflaton decay, shown for $m_\sigma = 100\GeV$. There is no perturbative decay for $m_\sigma = 10^{11}\GeV$.}
\end{figure}

We find that perturbative decay can occur before $H(t)\sim\Gamma$ (\fig{fig:pert}). Perturbative thermal decay is more efficient for larger $g$ and smaller $m_\sigma$. In fact, for large $m_\sigma$, this type of decay is never efficient before $H(t)\sim\Gamma$. Altering either $H_*$ or $\sigma_*$ does not change the results. Note that unlike the broad resonance, perturbative decay does not become blocked after it has begun.
\section{Results}\label{results}

\begin{figure}
\centering
\includegraphics[width=0.47\textwidth]{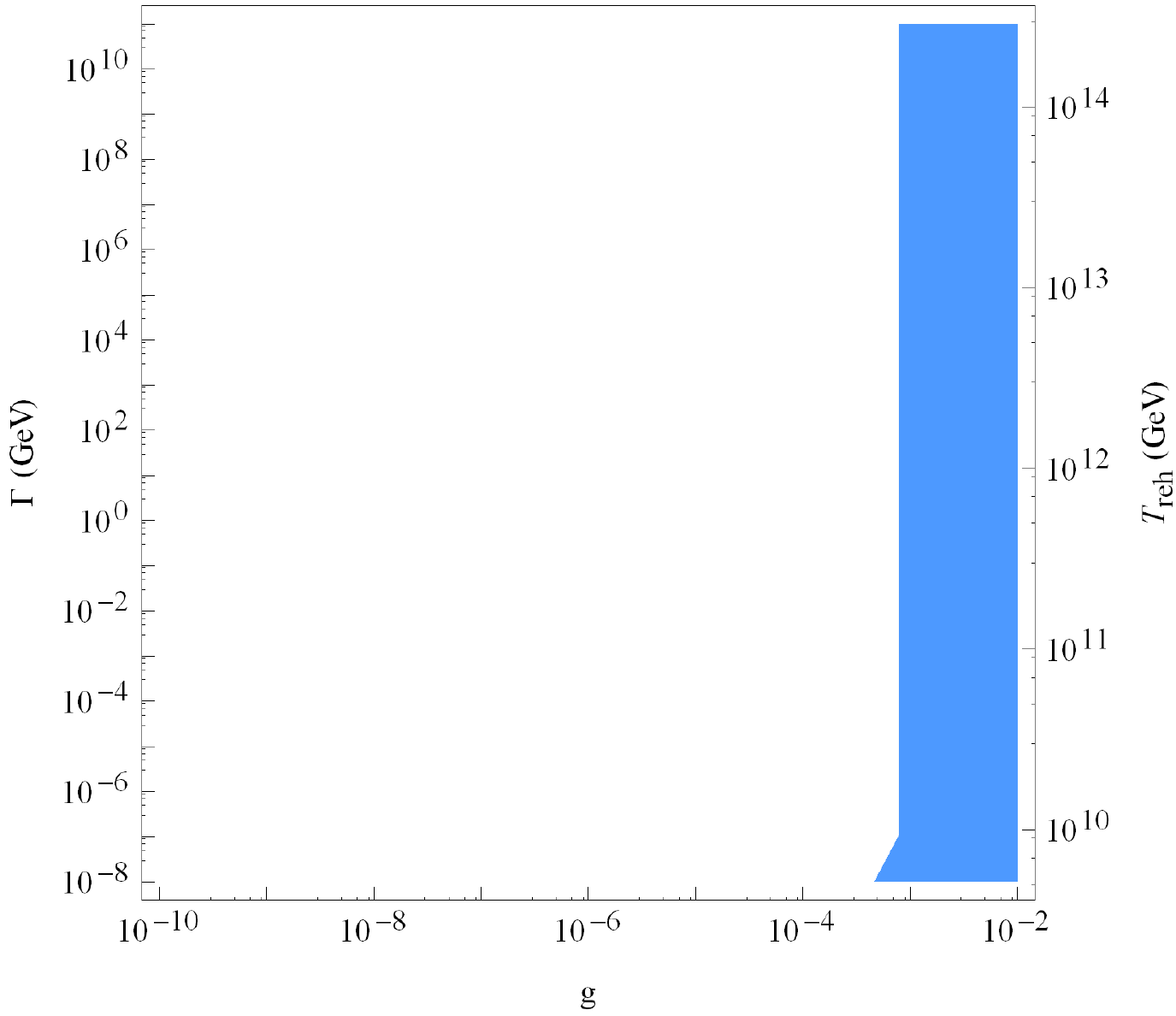} \qquad
\includegraphics[width=0.47\textwidth]{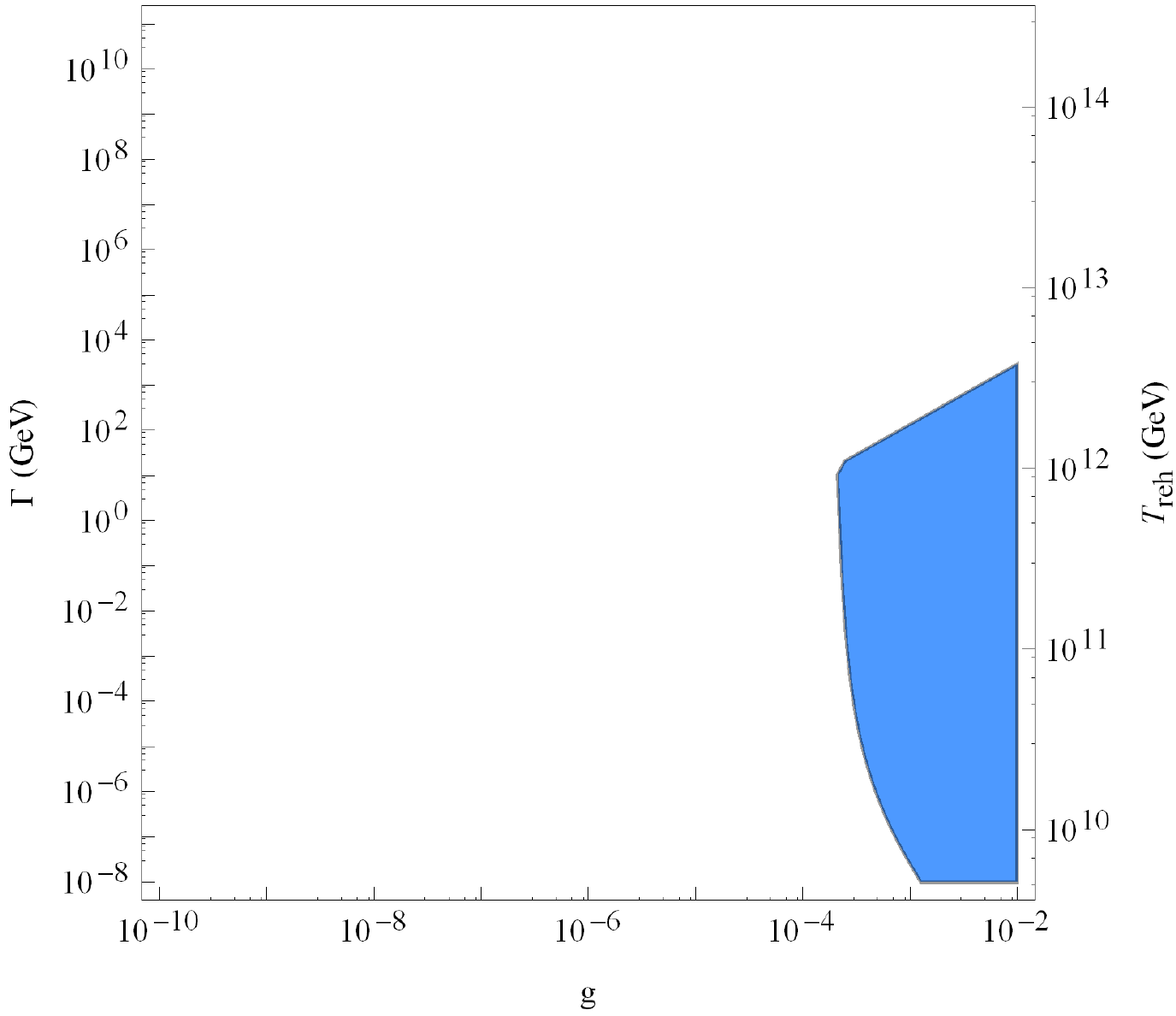}
\caption{\label{fig:results} Entire parameter space showing combined results from sections \ref{resgT}--\ref{pert}. Shaded region undergoes decay before $H \sim \Gamma$, by either broad resonance, narrow resonance or perturbative decay. Note the dependence on spectator mass: left, $m_\sigma = 100\GeV$; right, $m_\sigma = 10^{11}\GeV$. ($H_* = 10^{12}\GeV$, $\sigma_* = 10^{16}\GeV$.)}
\end{figure}

We collate the results from sections \ref{resgT}--\ref{pert} to give \fig{fig:results}. This shows the regions of the parameter space where the spectator decays before the inflaton. For both large and small $m_\sigma$, this occurs for $g\gtrsim (10^{-4} - 10^{-3})$. For small $m_\sigma$ there is no dependence on $\Gamma$. For $m_\sigma = 10^{11}\GeV$ the decay requires $\Gamma \lesssim 10^3\GeV$ (or equivalently a reheating temperature $T_{\mathrm{reh}}\lesssim 10^{12}\GeV$).

Varying $H_*$ only very slightly changes the parameter space, provided that the spectator remains light during inflation.  Decreasing $\sigma_*$ decreases the shaded region, although only by a small amount. The relative fraction of broad resonance compared to narrow resonance is also decreased. However, the qualitative results stand, that for small enough $\Gamma$ the decay is no longer blocked.

Thus, we see that the approximation of instant inflaton decay does not always well describe the dynamics of a realistic spectator decay, and predictions can be both quantitatively and qualitatively different. For large coupling ($g \gtrsim 10^{-4}$) and an (inflaton) reheating temperature of $\lesssim 10^{12}\GeV$, the phenomena of thermal blocking described in \cite{Enqvist:2012tc} does not occur (fixed $m_\sigma$, $H_*$ and $\sigma_*$). However, there does remain a large region of  parameter space where the qualitative features of the thermal blocking do not change.

These results have implications for the curvaton scenario. In a minimal scenario where the curvaton is coupled only to the standard model Higgs Boson, it is in principle possible to calculate the decay time of the curvaton as a function of model parameters. The time of curvaton decay determines the relative fraction of curvaton at decay. In combination with the initial field value, this gives both the magnitude of the curvature perturbation (known from observations) and the magnitude of the $\fnl$ parameter measuring non-Gaussianity (constrained from observations). It would then be possible to determine whether the model matches observations. The model of a curvaton coupled to the higgs has much more potential to be constrained than a more phenomenological curvaton model containing just an effective decay width.

Specifically, in this scenario the relative fraction of curvaton cannot start to grow until (i) the inflaton has decayed, and (ii) the spectator's potential is dominated by $m_\sigma^2 \sigma^2$. However, the curvaton {\em must} decay at least before BBN to avoid spoiling the predictions, and probably before dark matter freeze-out to avoid large isocurvature. Thus, a smaller $\Gamma$ means that there is less time for the relative fraction of curvaton to increase, and thus that a viable model is more difficult. In addition, a non-quadratic potential (i.e.\ case 3) causes the relative fraction of curvaton to either stay constant, or even decrease for some period. This also makes a viable curvaton model more difficult to achieve. A final point is that if decay occurs before $H(t)\sim\Gamma$ (shaded regions in \fig{fig:results}) then the curvaton model is immediately ruled out because the relative fraction of curvaton has no chance to grow.

Thus, in order to have a viable curvaton model, the effective decay rate of the inflaton should be sufficiently larger than that of the curvaton. A full study of this particular curvaton model is needed to determine the minimum allowed inflaton decay rate.

\section{Discussion and conclusions}\label{conc}

We have studied the non-perturbative, resonant decay of an oscillating spectator field coupled to standard model Higgs Bosons. The novel feature of our work is that we incorporated a more realistic assumption about inflaton decay, that it occurs with an effective width $\Gamma$ rather than instantaneously. Our focus was on the period with $H(t) > \Gamma$, i.e.\ while the universe is dominated by an inflaton oscillating in an effective quadratic potential. The thermal background produced by inflaton decay has a substantial impact on the spectator dynamics. In previous work on spectator non-perturbative decay, it was found that the large thermal mass gained by the higgs blocked the decay until some late time \cite{Enqvist:2012tc}.

In our case, the finite decay width of the inflaton has the effect of reducing the (inflaton) reheating temperature. Since the higgs thermal mass is given by $\gT T$, this means that the thermal mass is reduced for smaller $\Gamma$. This changes the higgs dispersion relation, which is decisive for resonant production of higgses and  the non-perturbative decay of the spectator. We also accounted for the induced decay of the oscillating spectator field due to the thermal background and found it to be important only in a limited region of the parameter space (large $g$ and small $m_\sigma$). Our new result is that for large enough $g$ and small enough $\Gamma$, the resonant decay is {\em not} blocked and resonant decay can occur quickly. This is bad for curvaton models, because the curvature perturbation will be too small. However, it is good for other spectator fields, which must decay early enough to avoid spoiling various cosmological predictions. Note that although the shaded region of parameter space (\fig{fig:results}) is small, it encompasses perhaps the most interesting region --- large couplings.

Our work also provides a qualitative confirmation that the phenomena of thermal blocking still exists even if the inflaton decay rate is slow, provided that the coupling is small enough ($g\lesssim 10^{-3}$). In the case where inflaton decay is not instantaneous, the technical results of this paper and of \cite{Enqvist:2012tc} should be combined in order to determine the dynamics of the system.

We observe a dependence on spectator mass; non-perturbative decay occurs predominantly for large spectator masses whereas perturbative decay is important for smaller masses. Although we have only presented results for the extreme cases of $100 \GeV$ and $10^{11}\GeV$, we expect a smooth transition between these regimes. It would be interesting to study the mass dependence of our results in more detail.

It would be an interesting and worthwhile project to study the implications of our results for the curvature perturbation. It is known from simulations that preheating effects do not tend to transfer all of the energy from the oscillating field \cite{Kofman:1997yn, Podolsky:2005bw}. The fraction of energy transferred should thus be determined by lattice calculations. That information would enable the effect on the perturbations to be calculated, provided that the final decay mechanism for the remainder of the oscillating field is known.

Preheating of the inflaton produces a background of stochastic gravitational waves \cite{Dufaux:2007pt,GarciaBellido:2007dg}. However, for the curvaton model these are not likely to be at frequencies and amplitudes accessible to currently-planned experiments unless the scale of inflation is particularly low \cite{Enqvist:2008be}. Gravitational waves from preheating in the curvaton model have been studied in \cite{Enqvist:2008be}, and it would be interesting to study this issue again in the context of the minimal curvaton-Higgs (MCH) model. We expect some effect on the spectrum as the inflaton decay width decreases. Other interesting effects could include anisotropy of the spectrum \cite{Bethke:2013aba}.

Our present work focuses on a specific minimal and semi-realistic model where the (singlet scalar) spectator couples only to the Higgs Boson of the standard model. Without adding any other new particles to the standard model, this is the only renormalisable coupling that is possible. However, the results do not rely on the scalar being the higgs (other than through the known value of $\gT$), only that the spectator's decay product is (fairly) strongly coupled to the thermal background.

In the case of the spectator being a {\em curvaton}, i.e.\ producing the curvature perturbation, we conclude that the slow decay of the inflaton makes a viable curvaton model more difficult. However, an advantage of the present scenario is that in principle the decay width of the curvaton can be calculated, rather than used as a free parameter. A forthcoming paper \cite{Enqvist:2013gwf} will detail our analysis of this particular curvaton model.

We have assumed that the dispersion relation of the produced higgs is instantly altered by thermal (or finite density) effects. It would be interesting to consider details of resonant production in conjunction with realistic, finite time thermalisation dynamics. Recent work on {\em perturbative} reheating in non-equilibrium quantum field theory includes \cite{Drewes:2013iaa}.

In summary, we have made progress towards a clear understanding of the complicated dynamics occurring in a `realistic' spectator field model with couplings to the standard model. We find that the assumption of instant inflaton decay leads to changes in the reheating temperature, expansion history and thus the decay time of the spectator. For some parameter values, this change is a huge qualitative shift, rather than an adjustment of parameters.

\section*{Acknowledgments}
KE and RL are respectively supported by the Academy of Finland grants 1218322 and  1263714; SR is supported by the V\"{a}is\"{a}l\"{a} foundation.

\appendix

\section{Higgs energy density from broad resonance}
In this section we calculate the energy density of the produced higgses, deriving \eq{finalrho_h} in the main text.

\numberwithin{equation}{section}
We need to estimate the integral
\begin{equation}
 I(j) = \frac{1}{2}\int_0^{\kmax^j} \ud k k^2 e^{2\pi\mu} \label{eq:integral},
\end{equation}
where $\mu \equiv \sum_{i = 1}^j \mu_k^i$ is an effective Floquet index. The majority of the particles will be created near $k=0$, so $\mu$ will have a maximum at the origin. Thus we use the saddle-point approximation and expand $\mu \simeq \mu_0 - \frac{1}{2}\left|\mu_0''\right| k^2$, with the prime denoting a derivative with respect to $k$ and
the subscript $0$ indicating that a quantity is evaluated at $k=0$.
Then the integral is just a Gaussian integral
\begin{eqnarray}
 I(j) & = &\frac{1}{2}\int_0^{\kmax^j} \ud k k^2 e^{2\pi\mu} \simeq \frac{1}{2}e^{2\pi \mu_0}\int_0^{\kmax^j} \ud k k^2 e^{-\pi|\mu_0''|k^2}
     \\ & \simeq & \frac{1}{4}e^{2\pi \mu_0}\int_{-\infty}^{\infty} \ud k k^2 e^{-\pi|\mu_0''|k^2}  \simeq \frac{e^{2\pi \mu_0}}{8\pi |\mu_0''|^{3/2}} \nonumber,
\end{eqnarray}
where $\mu_0$ and $\mu''_0$ depend on the case we are considering.

\subsection{Case 2}
For case 2, we find
\begin{eqnarray}
 2\pi\mu_0 & = & \sum_{i=1}^j \ln \left(1+2e^{-\sqrt{i/j_{\mathrm{block}}}}\right) \\
           & \simeq & \left(\ln 3 - \frac{4}{9}\sqrt{\frac{j}{j_{\mathrm{block}}}} + \frac{1}{18}\frac{j}{j_{\mathrm{block}}}\right)j
\end{eqnarray}
and
\begin{eqnarray}
 |\mu_0''|  & = & \frac{4}{3H_* g \sigstar}\sum_{i=1}^j\left[\frac{e^{-\sqrt{i/j_{\mathrm{block}}}}}{1 + 2e^{-\sqrt{i/j_{\mathrm{block}}}}}\right]a_i^{-1/2} \\
            & \simeq &  \frac{4}{9H_* g \sigstar}\sum_{i=1}^j\left(1-\frac{1}{3}\sqrt{i/j_{\mathrm{block}}}\right)a_i^{-1/2} \\
            & \simeq &  \left(\frac{16\msig}{81 \pi H_*}\right)^{1/3}\frac{j^{2/3}}{H_* g \sigstar} \left(1 - \frac{4}{21}\sqrt{\frac{j}{j_{\mathrm{block}}}}\right).
\end{eqnarray}
Thus, the integral in~\eqref{eq:integral} is
\begin{equation}
 I(j) \simeq \frac{9}{32  \sqrt{\pi}}\left(\frac{H_*}{\msig}\right)^2 \frac{(g\msig \sigstar)^{3/2}}{j\left(1 - \frac{4}{21}\sqrt{j/j_{\mathrm{block}}}\right)^{3/2}} e^{\left(\ln 3 - \frac{4}{9}\sqrt{\frac{j}{j_{\mathrm{block}}}} + \frac{1}{18}\frac{j}{j_{\mathrm{block}}}\right)j}
 \label{eq:integral_perpetualNR},
\end{equation}
and the energy density of the higgs particles produced is
\begin{equation}
 \rho_{\mathrm{h}} = \frac{4 m_h}{2\pi^2a_j^3}I(j) \simeq \frac{4 g \Sigma_j}{2\pi^2a_j^3}I(j) \simeq \frac{1}{4\sqrt{\pi^{11}}}\frac{(g\msig\sigstar)^{3/2}g\sigstar e^{\left(\ln 3 - \frac{4}{9}\sqrt{j/j_{\mathrm{block}}} + \frac{1}{18}j/j_{\mathrm{block}}\right)j}}{j^4 \left(1 - \frac{4}{21}\sqrt{j/j_{\mathrm{block}}}\right)^{3/2}}.
\end{equation}

\subsection{Case 3}
For case 3, we find
\begin{eqnarray}
 2\pi \mu_0 & = & \sum_{i=1}^j \ln\left(1+2e^{-\pi \kappa_{i,k=0}^2}\right) \\
            & \simeq & \sum_{i=1}^j\left[\ln 3 - \frac{2}{3}\left(\frac{a_i}{a_{\mathrm{block}}}\right)^{5/4} + \frac{1}{9}\left(\frac{a_i}{a_{\mathrm{block}}}\right)^{5/2}\right] \\
            &\simeq & 3^j - \frac{2}{3}f(5/2) + \frac{1}{9} f(5) \label{am0}
\end{eqnarray}
and
\begin{eqnarray}
 |\mu_0''|  & = & \frac{1.8}{g\sigma_* H_*}  \sum_{i=1}^j \left(\frac{e^{-\pi \kappa_{i,k=0}^2}}{1+2e^{-\pi \kappa_{i,k=0}^2}}\right) \\
            & \simeq & \frac{0.6}{g\sigma_* H_*} \sum_{i=1}^j\left[1 - \frac{1}{3}\left(\frac{a_i}{a_{\mathrm{block}}}\right)^{5/4} - \frac{1}{18}\left(\frac{a_i}{a_{\mathrm{block}}}\right)^{5/2}\right] \\
            &\simeq & \frac{0.6}{g\sigma_* H_*} \left( j - \frac{1}{3}f(5/2) - \frac{1}{18}f(5)\right). \label{amp}
\end{eqnarray}
Approximating the sums with integrals, we find
\begin{eqnarray}
 f(\alpha) \equiv \sum_{i=1}^j \left(\frac{a_i}{a_{\mathrm{block}}}\right)^{\alpha/2} & \simeq & \left(\frac{a_0}{a_{\mathrm{block}}}\right)^{\alpha/2}\left\{1.67^{\alpha} + \int_1^j \ud i \left[1+0.67(2i-1)\right]^{\alpha}\right\}
 \\
 & = & \frac{0.75}{(\alpha+1)}\left[\left(\frac{a_j}{a_{\mathrm{block}}}\right)^{\alpha/2}\left(\frac{a_j}{a_{0}^{1/2}}\right) + \left( 0.67(2\alpha+1) -1 \right)1.67^{\alpha}\left(\frac{a_0}{a_{\mathrm{block}}}\right)^{\alpha/2}\right] \nonumber,
\end{eqnarray}
where $a_0 = 1.24\tilde{\lambda}^{-1/3}$.
The energy density of the produced particles after the $j$th zero crossing is then
\begin{equation}
 \rho_{\mathrm{h}}^{j+1} \simeq \frac{4m_{h,j}}{2\pi^2 a_j^2}\int_0^{k_{\mathrm{max}}^j}\ud k k^2 \frac{1}{2}e^{2\pi\mu} \simeq \frac{5 g\sigstar e^{2\pi \mu_0}}{16\pi^2 \tilde{\lambda}^{1/3}a_j^3 |\mu_0''|^{3/2}}.
 \label{eq:rho_h_from_sigma_sig4_app}
\end{equation}

\section{Narrow resonance}
\label{sec:appendixB}
In this section we present the calculation for narrow resonance. The equation of motion for the higgs can be written as a Mathieu equation
\be{mathieu}
 \chi_k'' + \left[A_k - 2q\cos(2z)\right]\chi_k = 0,
 \ee
where
\be{defs}
 A_k = \frac{k^2/a^2 + g_T^2T^2}{\msig^2} +2q, \qquad z = \msig t + \frac{2\msig}{3H_*}, \qquad q = \frac{g^2\sigstar^2}{4\msig^2 z^2} . \nonumber
\ee
It is well known that the solutions of the Mathieu equation are amplified exponentially within narrow resonance bands characterised by $A_k^{(l)} \approx l^2$ and whose width
is given by $\Delta A_{k}^{(l)} \sim q^l$ where $l=1,2,3,... $. Since the modes evolve adiabatically we use `instantaneous' solutions of the Mathieu equation at each time and concentrate on the first band as it gives the largest contribution. The band is given by $A_k \sim 1 \pm q$, i.e.,
\begin{equation}
 \frac{k^2}{a^2 \msig^2} + \frac{g_T^2T^2}{\msig^2} = 1 - 2 q \pm q.
\end{equation}
When temperature is large no modes are inside the band but as it decreases the band becomes unblocked and resonance can occur. This happens for the middle of the band when
\begin{equation}
 g_T^2 T_{\mathrm{nr}}^2 = \msig^2(1-2\qnr).
\end{equation}
The Floquet index is given by~\cite{Enqvist:2012tc}
\begin{equation}
 \mu_k = \sqrt{\left(\frac{q}{2}\right)^2 - \left[\frac{(k^2/a^2 + g_T^2T^2)^{1/2}}{(1-2q)^{1/2}\msig} - 1\right]^2} \label{eq:mu_k}.
\end{equation}

\subsection{No inflaton decay}

First we consider the case with no inflaton decay, that is, after narrow resonance becomes unblocked the temperature scales as $T^2 = T_{nr}^2\left(\frac{\anr}{a}\right)^2 = g_T^{-2}(1-2\qnr)\msig^2\left(\frac{\anr}{a}\right)^2$. At the time of unblocking the resonance band extends from $k=0$ to $\Delta \knr \equiv \qnr^{1/2}\anr\msig$ in agreement with~\cite{Enqvist:2012tc};
however, in contrast to~\cite{Enqvist:2012tc}, we find that this quickly ceases to be the case as the center of the band moves towards higher (comoving) momenta and as the band becomes
narrower due to decreasing $q$. The modes which are initially inside the resonance band ($k<\Delta \knr$) will exit when
\begin{equation}
 k^2 = x_-^{2}\anr^2\msig^2\left[1 - 3\qnr x_-^{-3} - (1-2\qnr)x_-^{-2}\right] \qquad \mathrm{where} \qquad x_-\equiv \frac{a_-}{\anr} \label{eq:exit1}.
\end{equation}
A given mode spends within the band a time
\begin{equation}
 \Delta z \simeq \frac{3\qnr \znr }{4} \left[1+\left(\frac{k}{\Delta \knr}\right)^2\right] \ll \znr.
\end{equation}
Taking the maximum value for the magnitude of the Floquet index, $\mu_k \simeq q/2$, the occupation numbers are amplified
\begin{equation}
 n_k \simeq n_k^{(0)}e^{2\mu_k\Delta z} \simeq n_k^{(0)}e^{\frac{3\qnr^2\znr}{4}\left[1+\left(\frac{k}{\Delta \knr}\right)^2\right]},
\end{equation}
where $n_k^{(0)} \simeq (e^{g_T}-1)^{-1}$ is the initial occupation number given by the Bose-Einstein distribution (see \cite{Enqvist:2012tc}). In order to have significant production we
must have large $\qnr^2\znr$. Then the contribution of these modes to the energy density can be estimated as
\begin{eqnarray}
 \rho_{k<\Delta \knr} & = & \frac{4e^{\frac{3\qnr^2\znr }{4}}}{2\pi^2 a^3(e^{g_T}-1)}\int_{0}^{\Delta \knr}\ud k k^2 e^{\frac{3\qnr^2\znr }{4}\left(\frac{k}{\Delta \knr}\right)^2} \sqrt{(k/a)^2 + g_T^2T^2 + 2q \msig^2}
 \\
                   & \simeq &  \frac{2\msig(\Delta \knr)^3e^{\frac{3\qnr^2\znr }{4}}}{\pi^2 a^3(e^{g_T}-1)}\left(\frac{\anr}{a}\right)\int_{0}^{1}\ud u u^2 e^{\frac{3\qnr^2\znr }{4}u^2}
 \\
                        & \simeq &  \frac{4\msig^4e^{\frac{3\qnr^2\znr }{2}}}{3\pi^2(e^{g_T}-1)\qnr^{1/2}\znr }\left(\frac{\anr}{a}\right)^4 . \label{eq:initially_inside}
\end{eqnarray}
The modes which are initially outside of the resonance band ($k>\Delta \knr$) will enter the band when
\begin{equation}
 k^2 = x_+^{2}\anr^2\msig^2\left[1 - \qnr x_+^{-3} - (1-2\qnr)x_+^{-2}\right] \qquad \mathrm{where} \qquad x_+\equiv \frac{a_+}{\anr}. \label{eq:enter1}
\end{equation}
This gives
\begin{equation}
 x_+ = \frac{a_+}{\anr} \simeq  \left[(1-2\qnr) + \qnr\left(\frac{k}{\Delta \knr}\right)^2\right]^{1/2}  \simeq \left[(1-\qnr) + \qnr\left(\frac{k}{\Delta \knr}\right)^2\right]^{1/2}.
 \label{eq:cubic_solution}
\end{equation}
The time of exit is again given by~\eqref{eq:exit1}. Thus the modes will spend within the band the time
\begin{equation}
 \Delta z \simeq \frac{3q_+z_+}{2} = \frac{3\qnr \znr }{2}\left[(1-\qnr) + \qnr\left(\frac{k}{\Delta \knr}\right)^2\right]^{-3/4},
\end{equation}
and occupation numbers are amplified
\begin{equation}
 n_{k} \simeq n_{k}^{(0)}\mathrm{exp}\left\{\frac{3\qnr^2\znr }{2}\left[1-\qnr + \qnr\left(\frac{k}{\Delta \knr}\right)^2\right]^{-9/4}\right\}.
\end{equation}
Again we notice that $\qnr^2\znr$ must be large for efficient production. Thus the dominant contribution to the energy density integral will come from modes $k<\qnr^{-1}\Delta \knr$ (IR modes). Inverting~\eqref{eq:enter1} we get the mode the resonance band has reached in time $t$ which is the upper limit for the integral that can then be estimated as
 \begin{eqnarray}
 \rho_{k>\Delta \knr}(t) & = & \frac{4}{2\pi^2a^3}\int_{\Delta \knr}^{k_{\mathrm{cut}}(t)}\ud k k^2 \omega_k n_{k}^{(0)}\mathrm{exp}\left\{\frac{3\qnr^2\znr }{2}\left[1-\qnr + \qnr\left(\frac{k}{\Delta \knr}\right)^2\right]^{-9/4}\right\} \nonumber
 \\
                           & = & \frac{2\qnr^{3/2}\msig^3}{\pi^2}\left(\frac{\anr}{a}\right)^3\int_1^{k_{\mathrm{cut}}(t)/\Delta \knr}\ud u u^2 \omega_k(u) n_k^{(0)}(u) e^{\frac{3\qnr^2\znr }{2}(1+\qnr u^2-\qnr)^{-9/4}} \nonumber
 \\
 & \simeq & \frac{8\msig^4}{27\pi^2(e^{g_T}-1)}\frac{e^{\frac{3\qnr^2 \znr }{2}}}{\qnr^{3/2}\znr }\left(\frac{\anr}{a}\right)^4\left\{1-\left[\frac{(a/\anr)^2-1}{\qnr} +1\right]^{1/2}e^{-\frac{27\qnr^2\znr }{8}\left[(a/\anr)^2-1\right]}\right\} \nonumber.
\end{eqnarray}
Comparing this to~\eqref{eq:initially_inside} we see that the modes which are initially inside the resonance band contribute only a small fraction (of order $\qnr$) of the total energy density. We also note that the second term in the curly brackets quickly becomes negligible and we are left with the final energy density
\begin{equation}
 \rho_h \simeq  \frac{8\msig^4}{27\pi^2}\frac{e^{\frac{3\qnr^2 \znr }{2}}}{(e^{g_T}-1) \qnr^{3/2}\znr }\left(\frac{\anr}{a}\right)^4.
\end{equation}

\subsection{Including inflaton decay}
\label{appendixB2}
During the decay of the inflaton the temperature instead falls as $T_*a^{-3/8}$. The calculation proceeds as above and the modes initially inside the resonance band give a subdominant
contribution. The mode initially outside of the band will enter it when
\begin{equation}
 x_+ = \frac{a_+}{\anr} \simeq \frac{(1-4\qnr)}{2} + \frac{1}{2}\left[(1+4\qnr)^2 +5\qnr (u^2-1)\right]^{1/2}
\end{equation}
and spend within it the time
\begin{equation}
 \Delta z \simeq \frac{3q_+z_+}{2}\left[1-\frac{5}{8}\left(\frac{\anr}{a_+}\right)^{3/4} \right]^{-1}.
\end{equation}
Thus the modes get amplified according to
\begin{equation}
 n_k \simeq (e^{g_T} - 1)^{-1} \mathrm{exp}\left\{4\qnr^2\znr\left[1-\frac{115}{16}\frac{\qnr(k^2/\Delta \knr^2-1)}{(1+4\qnr)}\right]\right\}.
\end{equation}
As in the previous case large $\qnr^2\znr$ is needed for significant production. The energy density of produced particles evolves as
\begin{equation}
 \rho_{k>\Delta\knr} \simeq \frac{4\msig^4}{115(e^{g_T}-1)}\frac{e^{4\qnr^2\znr}}{\qnr^{3/2}\znr}\left(\frac{\anr}{a}\right)^{27/8}\left[1-u_{\mathrm{cut}}e^{-\frac{115\qnr^3\znr}{4}\left(u_{\mathrm{cut}}^2-1\right)}\right],
 \label{eq:rho_outside}
\end{equation}
where
\begin{equation}
 u_{\mathrm{cut}} \simeq \left[1+\frac{4}{5\qnr}\left(\frac{a}{\anr}-1\right)\left(\frac{a}{\anr}+4\qnr\right)\right]^{1/2},
\end{equation}
leaving
\begin{equation}
 \rho_h \simeq \rho_{k>\Delta\knr} \simeq \frac{4\msig^4}{115(e^{g_T}-1)}\frac{e^{4\qnr^2\znr}}{\qnr^{3/2}\znr}\left(\frac{\anr}{a}\right)^{27/8}
\end{equation}
once the production terminates.

\subsection{Transition from broad resonance}
\label{appendixB3}
For some values of the parameters the narrow resonance unblocking condition $g_TT<\msig$ becomes satisfied while the system is still in the broad resonance regime $q>1$. Then narrow resonance
begins once $q\approx 1$. In this case the resonance band is initially at $k \sim \anr \msig$, already far away from the origin (subscript `nr' still refers to the time of onset of narrow resonance which is now different than in the previous case). The modes now enter the band at
\begin{equation}
 x_+ = \frac{a}{\anr} \simeq \frac{k}{\anr\msig}\left(1+\frac{\qnr}{2(k/\anr\msig)^3} + \frac{g_T^2T_{\mathrm{nr}}^2}{2\msig^2(k/\anr\msig)^{3/4}}\right).
\end{equation}
Number densities are amplified according to
\begin{equation}
 n_k = n_k^{(0)} e^{2\mu_k\Delta z} \simeq n_k^{(0)}e^{\frac{3}{2}\qnr^2\znr x_+^{-9/2}}.
\end{equation}
In contrast to the previous case, now UV modes are produced so $\omega_k \simeq k/a$ and the initial occupation numbers are
\begin{eqnarray}
 n_k^{(0)} & = & \frac{1}{e^{\omega_k/T}-1} \simeq \frac{1}{e^{g_T u (g_TT_{\mathrm{nr}}/\msig)^{-1}x_+^{-5/8}}-1} \ll 1.
\end{eqnarray}
Thus the energy density of produced particles is
\begin{eqnarray}
 \rho_h \simeq  \frac{8\msig^4}{27\pi^2}\frac{e^{\frac{3}{2}\qnr^2\znr\left(1-\beta\right)}}{\qnr^2\znr\left(1+2\beta/9\right)}\left(\frac{a}{\anr}\right)^{-4}\left[1-\left(\frac{a}{\anr}\right)^3 e^{-\frac{27}{4}\left(1+\frac{2\beta}{3\alpha}\right)\left(\frac{a}{\anr}-1\right)}\right],
\end{eqnarray}
where $\beta \simeq 2\msig/3\qnr^2\znr T_{\mathrm{nr}}$. As before we must have large $\qnr^2\znr$ for significant production but now also $\beta < 1$ is required, otherwise there will be too
few particles initially for efficient decay of the spectator to occur.


\end{document}